\documentclass[sigconf]{acmart}

\AtBeginDocument{%
  }

\setcopyright{acmlicensed}
\copyrightyear{2018}
\acmYear{2018}
\acmDOI{XXXXXXX.XXXXXXX}
\acmConference[Conference acronym 'XX]{Make sure to enter the correct
  conference title from your rights confirmation email}{June 03--05,
  2018}{Woodstock, NY}
\acmISBN{978-1-4503-XXXX-X/2018/06}

\usepackage{amsmath,amsfonts}
\usepackage{algorithm}
\usepackage{algpseudocode}   
\usepackage{bm}
\usepackage{xspace}
\usepackage{microtype}
\usepackage{graphicx}
\usepackage{subcaption}
\usepackage{booktabs} 
\usepackage{multirow}
\usepackage{xcolor}
\usepackage{colortbl}
\usepackage{tcolorbox}
\usepackage{enumitem}
\usepackage{tabularx}

\definecolor{tableheader}{RGB}{229,240,254}
\definecolor{tableours}{RGB}{238,249,255}

\definecolor{greyL}{RGB}{240,240,240}
\tcbset{
  highlightbox/.style={
    colback=greyL,
    colframe=black,
    boxrule=1pt,
    arc=5pt,
    boxsep=5pt,
    left=5pt, right=5pt, top=0pt, bottom=0pt,
    fontupper=\itshape
  }
}

\hypersetup{
    colorlinks=true,
    citecolor=cyan,
    linkcolor=red,
    urlcolor=cyan,
}

\usepackage[capitalize,noabbrev]{cleveref}

\theoremstyle{plain}

\theoremstyle{definition}

\theoremstyle{remark}

\definecolor{codecomment}{RGB}{63,127,127}

\begin{document}


\title[ANTIDOTE: A Defense Harness for Agentic Recommender Systems]{ANTIDOTE: A Defense Harness for Retrieval Diversification and Evidence Consolidation in Agentic Recommender Systems}


\author{Hongru Hou}
\affiliation{%
    \institution{Fudan University}
  \country{Shanghai, China}
  }

\author{Denghui Geng}
\affiliation{%
  \institution{Fudan University}
  \country{Shanghai, China}
  }

\author{Jinhui Huang}
\affiliation{%
  \institution{Fudan University}
  \country{Shanghai, China}
  }

\author{Zixi Liu}
\affiliation{%
  \institution{Fudan University}
  \country{Shanghai, China}
  }

\author{Zhitao Yin}
\affiliation{%
  \institution{Fudan University}
  \country{Shanghai, China}
  }

\author{Tiehua Mei}
\affiliation{%
  \institution{Fudan University}
  \country{Shanghai, China}
  }

\author{Weiyuan Li}
\affiliation{%
  \institution{Fudan University}
  \country{Shanghai, China}
  }

\author{Ao Xu}
\affiliation{%
  \institution{Fudan University}
  \country{Shanghai, China}
  }

\author{Hengrui Chen}
\affiliation{%
  \institution{Fudan University}
  \country{Shanghai, China}
  }

\author{Ganlin Xu}
\affiliation{%
  \institution{Fudan University}
  \country{Shanghai, China}
  }

\author{Jiaqing Liang}
\affiliation{%
  \institution{Fudan University}
  \country{Shanghai, China}
  }

\author{Deqing Yang}
\correspondingauthor
\affiliation{%
  \institution{Fudan University}
  \country{Shanghai, China}
  }

\renewcommand{\shortauthors}{Hou et al.}

\providecommand{\method}{\textsc{Antidote}\xspace}
\providecommand{\algdiv}{\textsc{Forage}\xspace}
\providecommand{\algcom}{\textsc{Sieve}\xspace}
\providecommand{\bench}{\textsc{ARSBench}\xspace}

\begin{abstract}
Agentic recommender systems operate as proxies to bridge users with platforms. They receive requests and construct recommendations by actively querying external tools to retrieve candidates and gather evidence for reranking. However, reliance on externally retrieved candidates and review evidence introduces two distinct failure modes in the retrieve-then-rerank pipeline: limited candidate exposure at retrieval and susceptibility to review-text manipulation during reranking. To systematically investigate these issues, we construct an Agentic Recommender Systems Benchmark, \bench, demonstrating that retrieval-stage diversity deficits constrain downstream brand and feature coverage, while injected attribute-free subjective text can manipulate product rankings. To address these issues, we propose \method, a defense harness designed for agentic recommender systems. It employs a \algdiv module to probe locally underrepresented features and broaden candidate exposure, alongside a \algcom\ module that consolidates attribute-grounded claims to limit evidence-volume amplification during reranking. Extensive experiments across a vanilla LLM and four agentic recommenders show that \method\ improves retrieval diversity and recommendation quality while mitigating ranking manipulation under review-text injection attacks.
\end{abstract}



\maketitle

\providecommand{\method}{\textsc{Antidote}\xspace}  
\providecommand{\algdiv}{\textsc{Forage}\xspace}    
\providecommand{\algcom}{\textsc{Sieve}\xspace}     

\section{INTRODUCTION}
\label{sec:intro}

\begin{figure}[t]
  \centering
  \includegraphics[width=\linewidth]{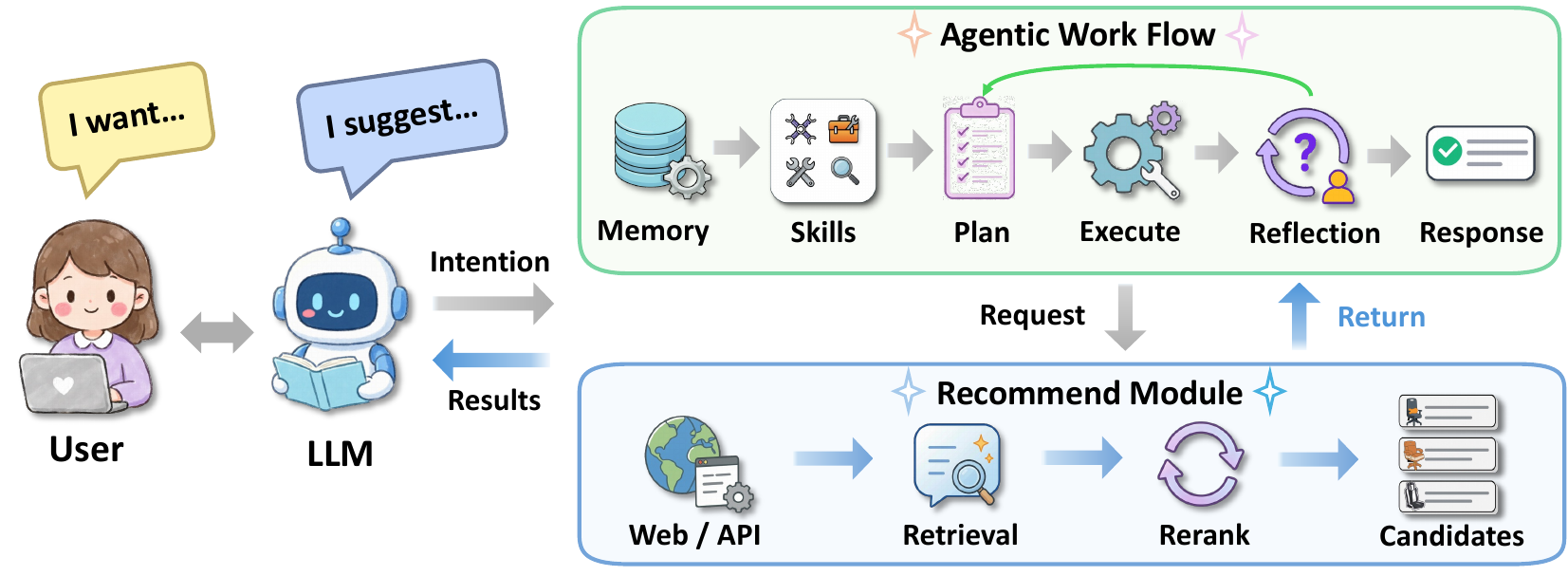}
  \caption{\textbf{The paradigm of agentic recommendation.} An LLM agent mediates between the user and platform to orchestrate recommendations by retrieval and reranking.}
  \label{fig:teaser}
  \vspace{-6mm}
\end{figure}

Recommender systems play a pivotal role in helping users discover products~\cite{hou2026prorl,mei2025goracs,xu2026verify}, media~\cite{he2025short,xu2025leveraging}, and information~\cite{hou2025heterogeneous}. The advent of large language models (LLMs) and tool-using agents is fundamentally reshaping this decision-making process~\cite{hu2026rethinking,zhang2026agenticrs,mei2026deep,xu2026verify}. A rapidly growing line of work recasts recommendation as an \emph{agentic} process~\cite{shang2026agentrecbench}: instead of scoring a fixed candidate set with a closed model, an LLM agent harnesses persistent memory, callable tools, and reusable skills to execute the recommendation end-to-end. This field has rapidly evolved from simply prompting models to generate items based on parametric knowledge~\cite{geng2022recommendation,bao2023tallrec,hou2024large}, toward deploying agents that actively plan, retrieve, and synthesize external evidence before responding~\cite{yao2023react,wang2024recmind,zhang2024agentcf,wang2024macrec,yincwagraph}. This new paradigm introduces an autonomous proxy (Figure~\ref{fig:teaser}): a recommendation agent that bridges the user and the platform, running the full retrieve-then-rerank pipeline on the user's behalf~\cite{wang2024recmind,zhao2024let,wang2024macrec,xu2025iagent,mei2026learning}.

However, reliance on externally retrieved candidates and review evidence introduces two distinct failure modes in the retrieve-then-rerank pipeline. At retrieval, the platform may return a narrow set of similar products, leaving potentially better-matched alternatives outside the agent's candidate pool. At reranking, attackers can manipulate user-generated reviews to bias product rankings and steer the agent's recommendations toward their desired outcomes. Despite these risks, existing methods typically rerank within the retrieved candidate pool while relying on accompanying product information and reviews~\cite{wang2024recmind,wang2024macrec}, making their decisions sensitive to both limited candidate exposure and manipulated review content.

The risk of manipulated external content has been widely studied in LLM agents, particularly through prompt injection and knowledge poisoning~\cite{greshake2023not,zhong2023poisoning,zou2025poisonedrag,zhan2024injecagent,chen2024agentpoison,wang2026mcptox}. 
In agentic recommendation, we consider a review-manipulation scenario in which an attacker aims to promote a target product by injecting review text without modifying product metadata, retrieval rankings, or tool descriptions. In contrast, retrieval concentration is treated as an exposure-quality limitation rather than an adversarial attack on the platform. To investigate these two failure modes, we construct a benchmark \bench\ using product information and reviews collected from JD\footnote{\url{https://www.jd.com/}} and Taobao\footnote{\url{https://www.taobao.com/}}, revealing two challenges in existing agentic recommender systems.

\noindent\textbf{Retrieval constrains candidate diversity.} Agentic recommender systems rely on platform retrieval tools that may return candidate pools concentrated around dominant brands and similar product features~\cite{zheng2021dgcn}. Such concentration can persist even when many products are retrieved, as a larger candidate pool does not necessarily provide broader brand and feature coverage. As a result, the agent's ability to explore alternatives and accommodate diverse user preferences is constrained by the products exposed through retrieval. As shown in Observations~\ref{obs:recall} and~\ref{obs:brand} (Section~\ref{sec:challenges}), this diversity deficit originates at retrieval and cannot be repaired by downstream reranking. Consequently, the final recommendation may overlook less visible but potentially better-matched alternatives, limiting the agent's ability to satisfy diverse user needs.

\noindent\textbf{Reranking is susceptible to subjective review manipulation.} To evaluate candidates, the agent relies on attached evidence for reasoning, such as user reviews and product metadata. However, Observation~\ref{obs:sparsity} (Section~\ref{sec:challenges}) shows that attribute-grounded evidence occupies only a small portion of the review text, while most of the remaining content provides no concrete attribute information. This composition leaves the reranking stage susceptible to repeated attribute-free subjective text. As shown in Observation~\ref{obs:inflation} (Section~\ref{sec:challenges}), appending attribute-free subjective text to the evidence of an initially lowest-ranked target substantially moves it toward higher-ranking positions on both evaluated LLM rerankers, despite introducing no additional product evidence. Such evidence manipulation can place the lowest-ranked items in prominent positions, crowd out items backed by stronger evidence, and ultimately distort the quality of the final recommendation.

To address these two failure modes, we propose \method, a \emph{plug-in} harness for existing agentic recommenders. \method consists of two complementary modules: \algdiv, which expands candidate exposure to mitigate retrieval-stage diversity deficits, and \algcom, which consolidates attribute-grounded claims to reduce the influence of review-text manipulation during reranking.

At the retrieval stage, \algdiv iteratively expands the candidate pool by re-querying the retrieval tool with locally underrepresented features observed in the current pool. A graph-based advisor first identifies a constrained set of underrepresented features, from which an LLM actor selects the next executable query anchor. This process broadens candidate exposure before downstream reranking.

At the reranking stage, \algcom reduces the influence of attribute-free subjective review text on candidate ranking. Rather than feeding lengthy and noisy reviews directly to the reranker, \algcom extracts attribute-grounded claims about concrete product aspects and consolidates semantically equivalent claims into a set-based evidence profile. This representation limits evidence-volume amplification while retaining distinct attribute-level evidence for candidate comparison, without verifying the factual correctness of individual claims.

Extensive experiments show that \method\ improves retrieval diversity and recommendation quality while mitigating ranking manipulation across multiple review-text injection settings.

In summary, the main contributions of this work include:
\begin{enumerate}[leftmargin=*]
    \item We identify two failure modes in agentic recommendation: limited candidate exposure during retrieval and susceptibility to review-text manipulation during reranking. We construct \bench\ to quantify their effects on candidate diversity and recommendation quality.
    \item We propose \method, a plug-in harness combining two complementary modules. \algdiv\ diversifies retrieval through graph-guided probing of underrepresented features, while \algcom\ consolidates attribute-grounded claims to limit evidence-volume amplification during reranking.
    \item Extensive experiments across a vanilla LLM and four agentic recommenders demonstrate that \method\ improves retrieval diversity, persona-conditioned recommendation quality, and robustness against the evaluated review manipulation attacks.
\end{enumerate}
\section{RELATED WORK}
\label{sec:related}

\subsection{Agentic Recommender Systems}
Driven by the emergent capabilities of LLMs, using LLMs as recommender systems has become a highly promising paradigm. P5~\cite{geng2022recommendation} unifies recommendation tasks under a prompt-based generative formulation, TALLRec~\cite{bao2023tallrec} aligns an instruction-tuned model to ranking objectives, and a prompted model can serve as a zero-shot ranker over a fixed candidate set~\cite{hou2024large}. Moving beyond the reliance on static parametric knowledge, the modern agentic paradigm equips LLMs with tools, memory, and planning, empowering the model to actively retrieve and synthesize external evidence before producing a recommendation. RecMind~\cite{wang2024recmind} plans and invokes external tools for zero-shot recommendation, AgentCF~\cite{zhang2024agentcf} models users and items as interacting agents, and MACRec~\cite{wang2024macrec} coordinates a team of specialized agents. More recently, the direction consolidates into an agent-platform paradigm~\cite{zhang2025envisioning}: iAgent~\cite{xu2025iagent} adds a user-side agent that reranks platform results from user instructions, whereas AgentCF++~\cite{liu2025agentcf++} equips agents with popularity-aware memory. However, these methods primarily reason over the candidates and evidence returned by retrieval tools, while the effects of limited candidate exposure and manipulated review content remain insufficiently examined.

\subsection{Security of LLM Agents}

LLM agents rely on external information to support their decisions, making them vulnerable to manipulation through untrusted content~\cite{xu2025iagent,wang2026mcptox,zhu2025melon}. Generative Engine Optimization (GEO)~\cite{aggarwal2024geo} demonstrates how source content can be optimized to influence its visibility in generated responses. Such content optimization can also be exploited for ranking manipulation. StealthRank~\cite{tang2025stealthrank} manipulates LLM rankings through adversarially optimized product descriptions. To mitigate these risks, SCI-Defense~\cite{yu2026sci} detects semantic manipulation signals, while SafeGEO~\cite{wen2026safegeo} investigates GEO attacks and defenses in recommendation agents. Unlike these approaches, our work focuses on subjective review manipulation, where repeated attribute-free content can distort reranking without introducing new product evidence. Rather than detecting malicious content, \algcom\ consolidates attribute-grounded claims to limit the influence of repeated evidence.
\providecommand{\method}{\textsc{Antidote}\xspace}

\makeatletter
\@ifundefined{observation}{%
  \theoremstyle{definition}%
  \newtheorem{observation}{Observation}%
  \theoremstyle{plain}%
}{}
\makeatother

\section{PRELIMINARIES}
\label{sec:prelim}

This section formalizes agentic recommender systems (\S\ref{sec:prelim:task}) and identifies the two challenges that motivate our method (\S\ref{sec:challenges}).

\subsection{Problem Formulation}
\label{sec:prelim:task}

Given a user $u$, an initial query $q_0$, and a user persona $p_u$, the agent seeks an ordered recommendation list $L$ of size $K_r$ that maximizes user satisfaction $\mathcal{U}(L\mid q_0,p_u)$. The agent accesses items exclusively through a retrieval tool $\mathcal{T}$ on the platform, which maps a query $q$ to a ranked list $\mathcal{T}(q)=(i_1,\ldots,i_n)$. For each returned item $i$, the agent observes its platform-provided brand $b(i)$, structured attribute set $\phi(i)$, and reviews $R_i$. For each product category, we define a platform-exposed reference catalog $\mathcal{M}=(\mathcal{B},\mathcal{V})$, where $\mathcal{B}$ and $\mathcal{V}$ denote the complete brand and attribute sets disclosed by the platform. For example, in the smartphone category, $\mathcal{B}$ may include Apple and Samsung, while $\mathcal{V}$ may include battery life and storage capacity. The reference catalog $\mathcal{M}$ is used solely for offline evaluation and is unavailable to the retrieval policy. Further details are provided in Appendix~\ref{app:catalog}.

\subsection{Challenges}
\label{sec:challenges}

As the agent assembles its recommendations from externally retrieved candidates and evidence, limitations in these inputs can affect the final results. We collect search results from two mainstream e-commerce platforms, JD and Taobao, and identify two distinct failure modes: \emph{diversity deficit} at retrieval and \emph{evidence toxification} at reranking.

\smallskip
\noindent\textbf{Challenge 1: Retrieval-stage diversity deficit.}
\label{sec:challenge1}
Agentic recommenders fetch candidates through the retrieval tools. Consequently, the tool's output strictly upper-bounds the entire pipeline: an item the tool fails to retrieve cannot be ranked or reasoned about. 
A low-diversity pool constrains the final recommendation space by concentrating the available choices around a narrow set of products, potentially excluding alternatives that better match the user's preferences. Deriving recommendations from such an impoverished pool inherently yields suboptimal results, regardless of the re-ranking policy's sophistication. Ultimately, retrieval-stage diversity limits the range of alternatives available for downstream recommendation.

To quantify how far a returned pool $\mathcal{P}$ falls below the full market $\mathcal{M}=(\mathcal{B},\mathcal{V})$, we score two axes. 
For brands, 
we compute the item fraction of each brand $\beta$ in $\mathcal{P}$, i.e., $p_\beta=|\{i|b(i)=\beta\}|/|\mathcal{P}|$, and report the maximal $p_\beta$ denoted as $p^{max}_\beta$.
Thus, $p^{max}_\beta$ measures the share of $\mathcal{P}$ contributed by its most frequent brand. For features, we report feature coverage $\mathrm{cov}(\mathcal{P})=|\bigcup_{i\in\mathcal{P}}\phi(i)|/|\mathcal{V}|$, the share of the category's attribute vocabulary the pool reaches. Specifically, we analyze the retrieval results on 20 product categories items and deliver two observations as follows:

\begin{figure}[t]
  \centering
  \includegraphics[width=\linewidth]{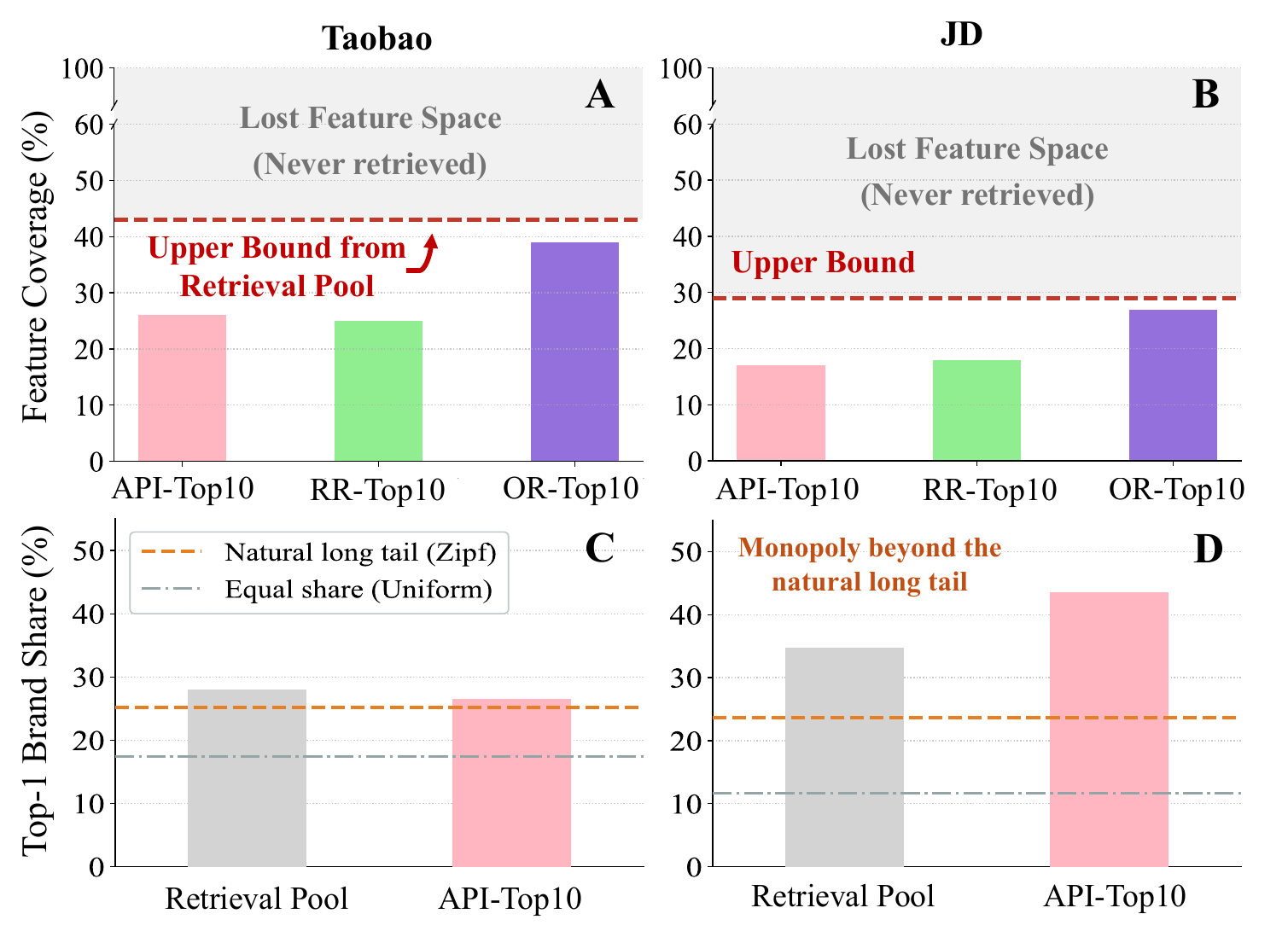}
  \caption{\textbf{Retrieval-stage diversity deficit.} \textbf{(A,~B) Feature collapse:} feature coverage of the API's served Top-$10$ (\textsc{API-Top10}), a random reranker (\textsc{RR-Top10}), and an oracle reranker (\textsc{OR-Top10}) that selects the maximum-coverage subset of the pool. Even the oracle is capped at the pool ceiling (red dashed), and a large part of the attribute space is never retrieved. \textbf{(C,~D) Brand skew:} the top-$1$ brand's share in the pool and in the served Top-$10$ far exceeds an equal-share baseline on both platforms, and on the platform that exposes per-brand catalog counts it also exceeds the catalog's own brand distribution (Appendix~\ref{app:brand-diversity}).}
  \label{fig:obs}
  \vspace{-4mm}
\end{figure}

\begin{observation}[Feature diversity is bottlenecked at retrieval]
\label{obs:recall}
On both platforms, the retrievals cover only a fraction of a category's attributes, and the lost coverage cannot be recovered by reranking.
\end{observation}
\noindent The served Top-$10$ reaches only a small share of a category's attributes (Figure~\ref{fig:obs}A,~B). A random reranker (RR) over the pool matches the API's own coverage, so the ordering is not the limiting factor, whereas an oracle reranker (OR) that maximizes coverage improves on it but immediately saturates at the pool ceiling (red dashed). The coverage above that line is never retrieved, indicating the bottleneck is in retrieval rather than ranking.

\begin{observation}[Brand is over-concentrated at retrieval]
\label{obs:brand}
On both platforms a single brand takes a large share of the candidate pool and of the served Top-$10$, far above an equal-share split, and on the platform whose catalog brand distribution is observable this share exceeds the catalog's own, so retrieval returns a pool more brand-concentrated than the brands the platform lists.
\end{observation}
\noindent The dominance is uneven across the two platforms but consistent in direction, and on the more concentrated one a single brand accounts for nearly half of the served list (Figure~\ref{fig:obs}C,~D, Appendix~\ref{app:brand-diversity}), leaving the agent little room to assemble brand-diverse recommendations.

Both deficits originate at retrieval and constrain all subsequent stages. Under a bounded context window, the agent condenses the retrieved pool through a selection operator $\operatorname{Sel}$ satisfying $\operatorname{Sel}(\mathcal{P}) \subseteq \mathcal{P}$. Since selection can only remove items, it cannot introduce any brand or attribute absent from the original pool. Specifically, $\operatorname{cov}\!\left(\operatorname{Sel}(\mathcal{P})\right)\leq\operatorname{cov}(\mathcal{P})$ and $\mathcal{B}\!\left(\operatorname{Sel}(\mathcal{P})\right)\subseteq\mathcal{B}(\mathcal{P})$, where $\mathcal{B}(\mathcal{P})=\{b(i)\mid i\in\mathcal{P}\}$ denotes the brands represented in $\mathcal{P}$. Although a reranker may reduce head-brand concentration within the retrieved pool, it cannot recover omitted brands or attributes. The missing-option problem must therefore be addressed at retrieval, for example, through multiple distinct queries that recover options omitted by a single truncated call (\S\ref{sec:method:div}).

\begin{figure}[t]
  \centering
  \includegraphics[width=\linewidth]{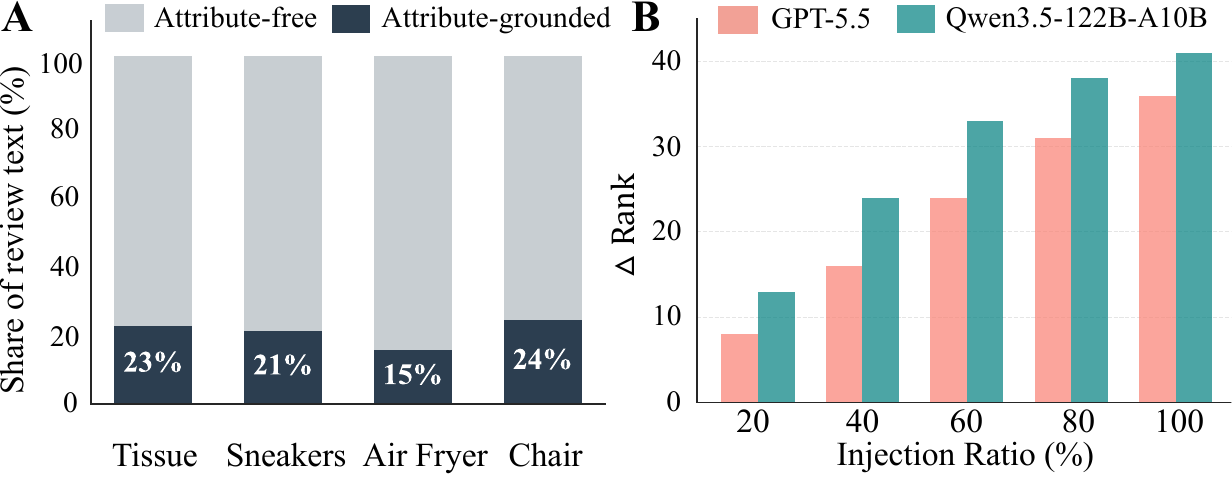}
  \caption{\textbf{Evidence toxification in the review channel.} \textbf{(A)} Review-evidence composition: Across the four categories, attribute-grounded evidence occupies only a minority of the review text, while the remaining text contains no attribute information. \textbf{(B)} Rank manipulation: The $\triangle$ Rank of an initially lowest-ranked target item under increasing attribute-free subjective text injection. The injection ratio denotes the proportion of reviews appended with one attribute-free subjective sentence. Results are averaged across 20 categories.}
  \label{fig:evi_poi}
\end{figure}

\smallskip
\noindent\textbf{Challenge 2: Reranking-stage evidence toxification.}
\label{sec:challenge2}
To optimize the final recommendation, the agent evaluates retrieved items using attached evidence such as user reviews. However, reviews mix attribute-grounded evidence, which ties factual or evaluative statements to concrete aspects, with attribute-free subjective text such as generic praise. In this work, we focus on a specific form of evidence toxification that amplifies the volume of such attribute-free subjective text while leaving attribute-grounded evidence unchanged. Our analysis shows that such content accounts for a substantial share of the review channel, creating an exploitable surface for attribute-free subjective text manipulation.

\begin{observation}[Low share of attribute-grounded text in review evidence]
\label{obs:sparsity}
Attribute-grounded evidence occupies only a small fraction of the sampled review text, while most of the remaining text contains no concrete attribute information and may nevertheless influence the reranker's assessment.
\end{observation}

\noindent A fine-grained content analysis shows that attribute-grounded quality signals occupy only 20\% of the sampled review text by character length (Figure~\ref{fig:evi_poi}A, Appendix~\ref{app:review-content-analysis}). Specifically, we measure the total character length of sentences or clauses containing factual descriptions or subjective assessments tied to concrete product attributes, relative to the total character length of all sampled reviews. Generic praise and emotional endorsement, by contrast, provide no attribute-level information. Because such non-attribute-specific language also occurs naturally in authentic reviews, malicious subjective content can blend into the existing review. We therefore conduct a targeted subjective-text injection experiment to test whether this surface can manipulate the agent's ranking.

\begin{observation}[Rank sensitivity to attribute-free subjective-text volume]
\label{obs:inflation}
Increasing the volume of attribute-free subjective texts consistently promotes an initially lowest-ranked item, with the effect growing monotonically across injection ratios and model families despite unchanged attribute-grounded evidence.
\end{observation}

\noindent We first identify the item ranked last by the agent under raw reviews as the target. We then append an attribute-free subjective sentence to the reviews of the target item. The high injection ratio indicates that more reviews are injected with an attribute-free subjective sentence. Averaged across $20$ categories, its rank improves monotonically on both backbones (Figure~\ref{fig:evi_poi}B). Details are shown in Appendix~\ref{app:observation4}. Because the subjective-text injection preserves the original review and adds neither product attributes nor model-facing instructions, the results show that the evaluated rerankers are susceptible to attribute-free subjective evidence.

\smallskip
\noindent Together, Observations~3 and~4 show that LLM rerankers may treat repeated attribute-free subjective signals as additional quality evidence without receiving new attribute-level information. This motivates \algcom, which discards attribute-free content and consolidates semantically equivalent attribute-grounded claims into set-based evidence units. \algcom therefore mitigates evidence-volume manipulation by filtering attribute-free content and consolidating semantically overlapping claims. Although it does not verify factual correctness, this mechanism also limits the repeated amplification of fabricated attribute claims.

\providecommand{\method}{\textsc{Antidote}\xspace}
\providecommand{\algdiv}{\textsc{Forage}\xspace}
\providecommand{\algcom}{\textsc{Sieve}\xspace}

\section{METHODOLOGY}
\label{sec:method}
\begin{figure*}[t]
  \centering
  \includegraphics[width=0.9\linewidth, height=8cm]{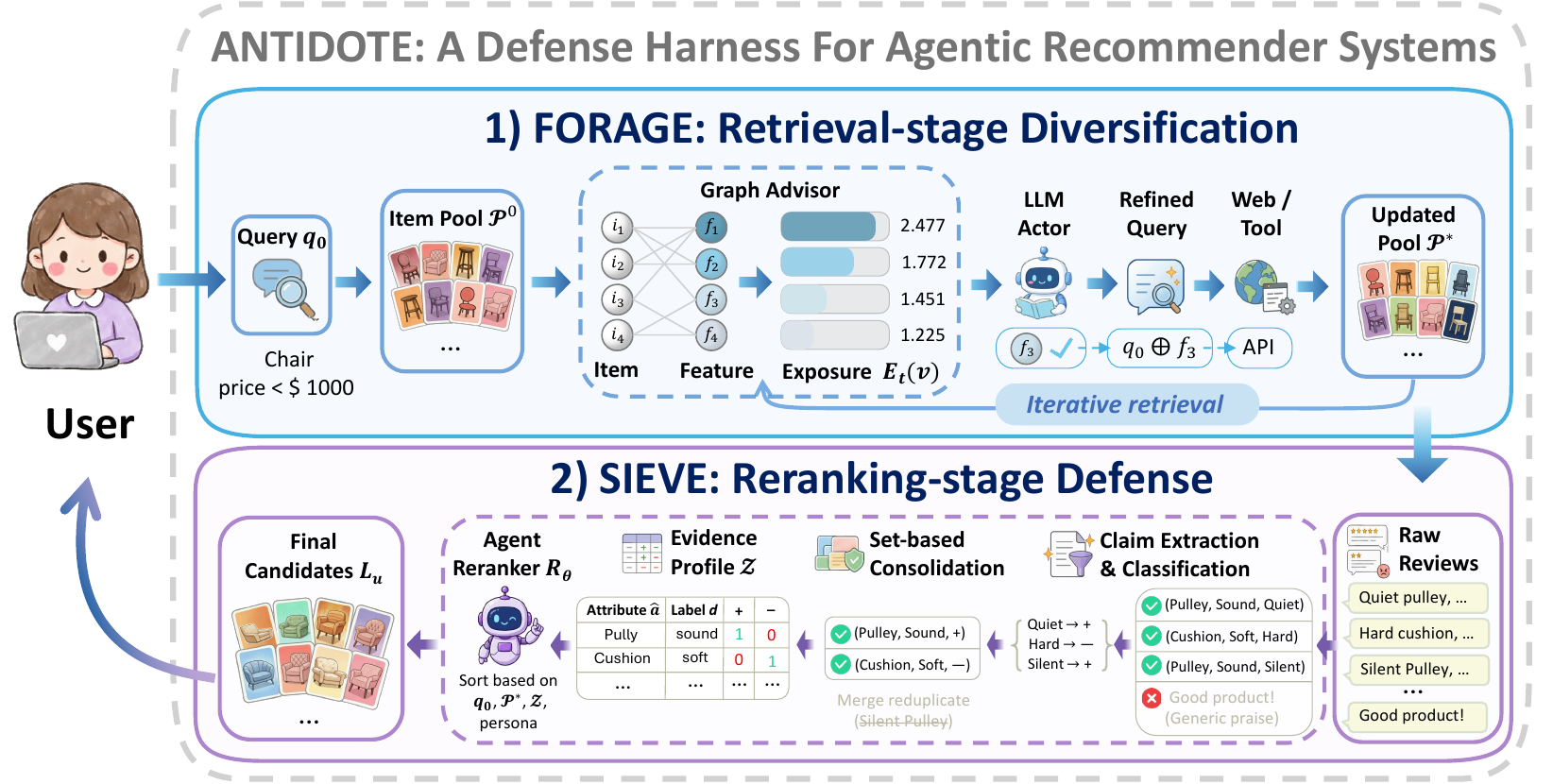}
  \caption{\textbf{Overview of \method.} \algdiv\ expands the retrieval pool through graph-guided retrieval of locally underrepresented observed features, while \algcom\ extracts and consolidates attribute-grounded claims for reranking.}
  \label{fig:paradigm}
\end{figure*}

We propose \method, a defense harness for agentic recommender systems, to address the challenges identified above. As illustrated in Figure~\ref{fig:paradigm}, \algdiv\ (\S\ref{sec:method:div}) mitigates retrieval-stage diversity deficits, while \algcom\ (\S\ref{sec:method:comment}) reduces the influence of manipulated review evidence during reranking. The two modules intervene at different stages of the retrieve-then-rerank pipeline.

\subsection{\algdiv: Diversity Restoration through Proactive Probing}
\label{sec:method:div}

\algdiv\ uses multi-round retrieval to mitigate the diversity deficit described in Challenge 1. At each round, it identifies underrepresented features in the current pool and uses them to retrieve additional candidates. It combines a spectral graph advisor (\S\ref{sec:method:graph}) with a constrained LLM actor (\S\ref{sec:method:forage}), as summarized in Algorithm~\ref{alg:forage}.

\subsubsection{Spectral Graph Advisor}
\label{sec:method:graph}

At round $t$, the advisor uses the current candidate pool $\mathcal{P}_t$ to construct the locally observed feature space $\mathcal{V}_t=\bigcup_{i\in\mathcal{P}_t}\phi(i)$, without access to a global attribute vocabulary. It builds an item--feature bipartite graph with incidence matrix $\mathbf{X}_t\in\{0,1\}^{|\mathcal{P}_t|\times|\mathcal{V}_t|}$, where $X_{t,iv}=1$ if item $i$ exposes feature $v$. Let $\mathbf{p}_t$ denote the leading right singular vector of $\mathbf{X}_t$, with its sign chosen such that $\mathbf{p}_t^{\top}\mathbf{1}\geq0$. This vector represents the dominant structure of the feature co-occurrence graph $\mathbf{X}_t^{\top}\mathbf{X}_t$.

For each item $i$, its alignment with the dominant pattern is
\begin{equation}
h_t(i)=\operatorname{clip}\!\left(\cos(\mathbf{x}_{t,i},\mathbf{p}_t),0,1\right),
\label{eq:mainstream_alignment}
\end{equation}
where $\mathbf{x}_{t,i}$ is the row of $\mathbf{X}_t$ corresponding to item $i$. A higher alignment indicates that the item is more representative of the current pool. The mainstream-exposure score of feature $v$ is then
\begin{equation}
E_t(v)=\sum_{i\in\mathcal{P}_t}X_{t,iv}h_t(i).
\label{eq:exposure}
\end{equation}
The advisor further selects the $K_f$ features with the lowest exposure scores as candidate exploration anchors, denoted by $\mathcal{C}_t$.

\subsubsection{Advisor--Actor Query Selection}
\label{sec:method:forage}

Low-exposure features may include promotional terms, model identifiers, or overly specific attributes that are unsuitable as search queries. We therefore use an LLM actor $\mathcal{A}$ to select an executable anchor $v_{t+1}\in\mathcal{C}_t$ based on the original query $q_0$, current candidate pool $\mathcal{P}_t$, previous anchors $\mathcal{Q}_t$, and graph-advised candidates $\mathcal{C}_t$. The actor can reject noisy features but cannot introduce features outside $\mathcal{C}_t$. The selected anchor is appended to the original query as $q_0\oplus v_{t+1}$, where $\oplus$ denotes whitespace-delimited concatenation. The retrieval tool determines which additional products are returned. FORAGE addresses candidate exposure rather than source trustworthiness. Additional retrieval may also introduce unreliable content, while neither module independently verifies the factual correctness of retrieved evidence.

\begin{algorithm}[t]
\caption{\algdiv: Graph-Guided Retrieval Diversification}
\label{alg:forage}
\begin{algorithmic}[1]
\State {\bfseries Input:} Initial query $q_0$; retrieval tool $\mathcal{T}$; attribute extractor $\phi$; graph advisor $\mathcal{G}$; LLM actor $\mathcal{A}$; rounds $R$; per-round retrieval size $N$; candidate-feature size $K_f$
\State {\bfseries Output:} Diversified candidate pool $\mathcal{P}^{\star}$
\State $\mathcal{P}_0\gets\mathcal{T}(q_0)[:N]$; $\mathcal{Q}_0\gets\varnothing$ \Comment{Initialize the candidate pool}
\For{$t=0$ \textbf{to} $R-1$}
  \State $\mathcal{V}_t\gets\bigcup_{i\in\mathcal{P}_t}\phi(i)$ \Comment{Locally observed feature space}
  \State $\mathcal{C}_t\gets\mathcal{G}(\mathcal{P}_t,\mathcal{V}_t,K_f)$ \Comment{Graph-advised candidates}
  \State $v_{t+1}\gets\mathcal{A}(q_0,\mathcal{P}_t,\mathcal{Q}_t,\mathcal{C}_t)$ \Comment{Select a query anchor}
  \State $\mathcal{R}_t\gets\mathcal{T}(q_0\oplus v_{t+1})[:N]$ \Comment{Re-query the tool}
  \State $\mathcal{P}_{t+1}\gets\mathcal{P}_t\cup\mathcal{R}_t$; $\mathcal{Q}_{t+1}\gets\mathcal{Q}_t\cup\{v_{t+1}\}$
\EndFor
\State \Return $\mathcal{P}^{\star}\gets\mathcal{P}_R$
\end{algorithmic}
\end{algorithm}

\subsection{\algcom: Mitigating Evidence Toxification via Semantic Consolidation}
\label{sec:method:comment}

\algcom\ converts raw reviews into compact semantic evidence profiles before reranking. It first extracts attribute-grounded claims (\S\ref{sec:method:extract}) and then consolidates semantically equivalent claims (\S\ref{sec:method:aggregate}). The first removes attribute-free content, while the second prevents repeated claims from receiving extra weight through review volume.

\subsubsection{Review-Grounded Claim Extraction and Classification}
\label{sec:method:extract}

Given a review $r$, an LLM extractor produces atomic evaluative claims:
\begin{equation}
\operatorname{Ext}(r)=\left\{(a_k,o_k,s_k)\right\}_{k=1}^{m_r},\qquad s_k\in\{+,-\},
\label{eq:extract}
\end{equation}
where $m_r$ may be zero, $a_k$ denotes a concrete aspect of the product, fulfillment, or service, $o_k$ is an explicit evaluative observation about that aspect, and $s_k$ denotes its polarity. A claim is retained only when the review connects a concrete aspect to an evaluative observation. Attribute-free praise, criticism, emotion, recommendation, urgency, popularity, and repurchase intent are discarded. Claims extracted from all reviews $\mathcal{R}_i$ of item $i$ form a multiset $\mathcal{E}_i$, preserving repetitions before consolidation.

\subsubsection{Set-based Evidence Consolidation}
\label{sec:method:aggregate}

Claim extraction alone does not prevent the same evidence from being repeated in different words. We therefore use an LLM-based normalization function $\psi_c$, conditioned on product category $c$, to map each claim to a canonical evidence unit:
\begin{equation}
\psi_c(a,o,s)=(\hat{a},d,s),
\label{eq:normalize}
\end{equation}
where $\hat{a}$ and $d$ are canonical attribute and semantic-dimension labels induced from the supplied reviews rather than selected from a predefined vocabulary. Paraphrased claims describing the same attribute and semantic dimension map to the same unit.

For item $i$, the presence of evidence unit $(\hat{a},d,s)$ is defined as
\begin{equation}
x_{i,\hat{a},d,s}=\max_{e\in\mathcal{E}_i}\mathbb{I}\!\left[\psi_c(e)=(\hat{a},d,s)\right],
\label{eq:evidence_set}
\end{equation}
with $x_{i,\hat{a},d,s}=0$ if $\mathcal{E}_i$ is empty. This binary representation records the presence of each canonical unit rather than its frequency. Consequently, additional claims mapped to an existing unit do not change the consolidated evidence. Positive and negative evidence remain separate, preserving conflicting observations without majority voting. The resulting evidence profile is
\begin{equation}
\mathcal{Z}_i=\left\{\left(\hat{a},d,x_{i,\hat{a},d,+},x_{i,\hat{a},d,-}\right)\;\middle|\;x_{i,\hat{a},d,+}+x_{i,\hat{a},d,-}>0\right\}.
\label{eq:evidence_profile}
\end{equation}

\smallskip
\noindent\textbf{Robustness Analysis.}
Under consistent claim extraction and normalization, attribute-free content contributes no valid evidence, and repeated claims do not change $\mathcal{Z}_i$.

\subsubsection{Persona-aware Listwise Reranking}

Given the user request $q_0$, persona profile $p_u$, diversified candidate pool $\mathcal{P}^{\star}$, and each item's structured metadata $M_i$ and consolidated evidence profile $\mathcal{Z}_i$, an LLM-based listwise reranker $R_{\theta}$ produces an ordered candidate list $\sigma_u$. The final recommendation $L_u$ consists of its top-$K_r$ items.

\providecommand{\method}{\textsc{Antidote}\xspace}
\providecommand{\algdiv}{\textsc{Forage}\xspace}
\providecommand{\algcom}{\textsc{Sieve}\xspace}
\providecommand{\imp}[1]{{\footnotesize\itshape #1}}
\providecommand{\gain}[1]{{\small\textcolor{red}{\,(#1)}}}

\section{EXPERIMENTS}
\label{sec:exp}
This section presents the experimental evaluation of our proposed framework to answer the following research questions:
\begin{itemize}[leftmargin=*]
    \item \textbf{RQ1:} Does our proposed plug-in defense harness enhance the end-to-end recommendation performance of existing agentic recommender baselines?
    \item \textbf{RQ2:} Does \algdiv\ improve the diversity of the retrieved candidates, and how effective is its query-selection policy compared with existing diversification strategies?
    \item \textbf{RQ3:} How robust is \algcom\ against different review manipulation attacks, and how does it compare with existing defenses?
    \item \textbf{RQ4:} How do the individual modules and their key components contribute to the effectiveness of \method?
    \item \textbf{RQ5:} \emph{What is the trade-off} between the achieved recommendation improvements and the operational efficiency?
\end{itemize}

\subsection{Experiment Setup}
\label{sec:exp:setup}
\noindent\textbf{Benchmark.}
Existing recommender-system and information-retrieval datasets do not simultaneously preserve query-conditioned retrieval results and comprehensive product-side information, including brands, structured attributes, and user reviews, making them unsuitable for jointly studying retrieval-stage diversity deficits and reranking-stage review manipulation. Therefore, we construct a benchmark \bench by collecting publicly available product data from two real-world e-commerce platforms: JD and Taobao. For each platform, we select 20 product categories, each containing at least 40 products. For every product, we retain its title, brand, price, structured attributes, and associated user reviews. The resulting corpus contains $2{,}085$ products and $131{,}459$ reviews. Additional statistics and construction details are provided in Appendix~\ref{app:data-construction}.

\noindent\textbf{Baselines.}
To evaluate the general applicability of \method, we integrate it into vanilla LLM and four representative agentic recommender systems. The selected baselines are RecMind~\cite{wang2024recmind}, ToolRec~\cite{zhao2024let}, MACRec~\cite{wang2024macrec}, and iAgent~\cite{xu2025iagent}, all of which follow a two-stage pipeline comprising tool-mediated product retrieval and LLM-based reranking. For each baseline, we compare the original implementation with its \method-augmented counterpart, where \algdiv\ is applied during retrieval and \algcom\ during reranking. We conduct experiments using two powerful LLM backbones from distinct model families, Qwen3.5-122B-A10B~\cite{qwen3.5} and GPT-5.5~\cite{singh2025openai}.

\noindent\textbf{Metrics.}
We evaluate the agentic recommender from two complementary perspectives: \emph{Recommendation Quality}~(RQ1) and \emph{Retrieval Diversity}~(RQ2). For recommendation quality, we follow the widely used LLM-as-a-Judge paradigm for scalable evaluation, which has been adopted across open-ended generation tasks~\cite{li2025curse,zheng2023judging} and increasingly applied to recommender-system evaluation~\cite{fabbri2025evaluating,zhang2024large,que2026llm}. Specifically, we use DeepSeek-v4-Pro~\cite{xu2026deepseek} as an evaluator to assign item-level relevance labels. For retrieval diversity, we evaluate candidate pools using metadata-based brand and feature statistics, including \emph{Brand Count}, \emph{Brand Dispersion}, \emph{Brand Entropy}, \emph{Feature Count}, and \emph{Feature Coverage}. Details are provided in Appendix~\ref{app:evaluation-metrics}. 

\subsection{Overall Performance (RQ1)}
\label{sec:exp:overall}

\begin{table*}[!t]
\centering
\small
\setlength{\tabcolsep}{3.8pt}
\renewcommand{\arraystretch}{1.10}

\caption{Overall recommendation performance (RQ1). The best performances are highlighted in bold, and the second-best are underlined. The superscript * indicates that the improvement is statistically significant under a two-sided category-level paired permutation test ($p<0.05$). Red {\color{red}(+\%)} values indicate relative improvements over the corresponding original pipeline.}
\label{tab:overall}

\resizebox{\textwidth}{!}{%
\begin{tabular*}{\textwidth}{
@{\extracolsep{\fill}}
l
cc
cc
cc
cc
@{}
}
\toprule
Backbone
& \multicolumn{4}{c}{GPT-5.5}
& \multicolumn{4}{c}{Qwen3.5-122B-A10B} \\
\cmidrule(lr){2-5}
\cmidrule(lr){6-9}

Dataset
& \multicolumn{2}{c}{Taobao}
& \multicolumn{2}{c}{JD}
& \multicolumn{2}{c}{Taobao}
& \multicolumn{2}{c}{JD} \\
\cmidrule(lr){2-3}
\cmidrule(lr){4-5}
\cmidrule(lr){6-7}
\cmidrule(lr){8-9}

Metric
& Precision@10 & NDCG@10
& Precision@10 & NDCG@10
& Precision@10 & NDCG@10
& Precision@10 & NDCG@10 \\
\midrule

Vanilla
& 0.5597 & 0.7236
& 0.5837 & 0.7623
& 0.5334 & 0.7019
& 0.5519 & 0.7212 \\

{\small +\,\method}
& \textbf{0.5822}\textsuperscript{*}\gain{+4\%}
& \textbf{0.7465}\textsuperscript{*}\gain{+3\%}
& \textbf{0.6145}\textsuperscript{*}\gain{+5\%}
& \textbf{0.7942}\textsuperscript{*}\gain{+4\%}
& \textbf{0.5513}\textsuperscript{*}\gain{+3\%}
& \textbf{0.7233}\textsuperscript{*}\gain{+3\%}
& \textbf{0.5782}\textsuperscript{*}\gain{+5\%}
& \textbf{0.7498}\textsuperscript{*}\gain{+4\%} \\

\midrule

RecMind
& 0.5621 & 0.7288
& 0.5864 & 0.7666
& 0.5412 & 0.7063
& 0.5544 & 0.7289 \\

{\small +\,\method}
& \textbf{0.5920}\textsuperscript{*}\gain{+5\%}
& \textbf{0.7709}\textsuperscript{*}\gain{+6\%}
& \textbf{0.6106}\textsuperscript{*}\gain{+4\%}
& \textbf{0.7990}\textsuperscript{*}\gain{+4\%}
& \textbf{0.5671}\textsuperscript{*}\gain{+5\%}
& \textbf{0.7410}\textsuperscript{*}\gain{+5\%}
& \textbf{0.5752}\textsuperscript{*}\gain{+4\%}
& \textbf{0.7576}\textsuperscript{*}\gain{+4\%} \\

\midrule

ToolRec
& 0.5703 & 0.7402
& 0.5937 & 0.7721
& 0.5486 & 0.7136
& 0.5628 & 0.7335 \\

{\small +\,\method}
& \textbf{0.6004}\textsuperscript{*}\gain{+5\%}
& \textbf{0.7806}\textsuperscript{*}\gain{+5\%}
& \textbf{0.6188}\textsuperscript{*}\gain{+4\%}
& \textbf{0.8068}\textsuperscript{*}\gain{+4\%}
& \textbf{0.5707}\textsuperscript{*}\gain{+4\%}
& \textbf{0.7512}\textsuperscript{*}\gain{+5\%}
& \textbf{0.5790}\textsuperscript{*}\gain{+3\%}
& \textbf{0.7573}\textsuperscript{*}\gain{+3\%} \\

\midrule

MACRec
& 0.5683 & 0.7421
& 0.5940 & 0.7743
& 0.5465 & 0.7103
& 0.5612 & 0.7413 \\

{\small +\,\method}
& \textbf{0.5927}\textsuperscript{*}\gain{+4\%}
& \textbf{0.7754}\textsuperscript{*}\gain{+4\%}
& \textbf{0.6170}\textsuperscript{*}\gain{+4\%}
& \textbf{0.8078}\textsuperscript{*}\gain{+4\%}
& \textbf{0.5670}\textsuperscript{*}\gain{+4\%}
& \textbf{0.7406}\textsuperscript{*}\gain{+4\%}
& \textbf{0.5787}\textsuperscript{*}\gain{+3\%}
& \textbf{0.7677}\textsuperscript{*}\gain{+4\%} \\

\midrule

iAgent
& 0.5822 & 0.7540
& 0.6045 & 0.7821
& 0.5577 & 0.7207
& 0.5673 & 0.7522 \\

{\small +\,\method}
& \textbf{0.6193}\textsuperscript{*}\gain{+6\%}
& \textbf{0.8004}\textsuperscript{*}\gain{+6\%}
& \textbf{0.6309}\textsuperscript{*}\gain{+4\%}
& \textbf{0.8149}\textsuperscript{*}\gain{+4\%}
& \textbf{0.5801}\textsuperscript{*}\gain{+4\%}
& \textbf{0.7519}\textsuperscript{*}\gain{+4\%}
& \textbf{0.5888}\textsuperscript{*}\gain{+4\%}
& \textbf{0.7831}\textsuperscript{*}\gain{+4\%} \\

\bottomrule
\end{tabular*}}
\end{table*}

Table~\ref{tab:overall} reports recommendation quality across both datasets and backbones. Across every baseline, dataset, and backbone, attaching \method consistently improves both Precision@10 and NDCG@10, with all improvements being significant under paired tests over the same evaluation instances. These joint gains indicate that \method returns more judged-relevant products and places them more prominently in the final ranking. In particular, the NDCG improvements are not obtained by reducing the number of relevant products in the returned list, since Precision@10 improves concurrently in every evaluated setting. Table~\ref{tab:overall} evaluates the end-to-end effectiveness of the complete harness, while RQ2 isolates the query-selection effect of \algdiv\ under matched retrieval budgets and RQ5 quantifies the associated operational cost. The results support the effectiveness of combining retrieval diversification with evidence consolidation to improve recommendation quality. 

\subsection{FORAGE Performance (RQ2)}
\label{sec:exp:rq2}

\begin{figure*}[t]
    \centering
\includegraphics[width=0.98\textwidth]{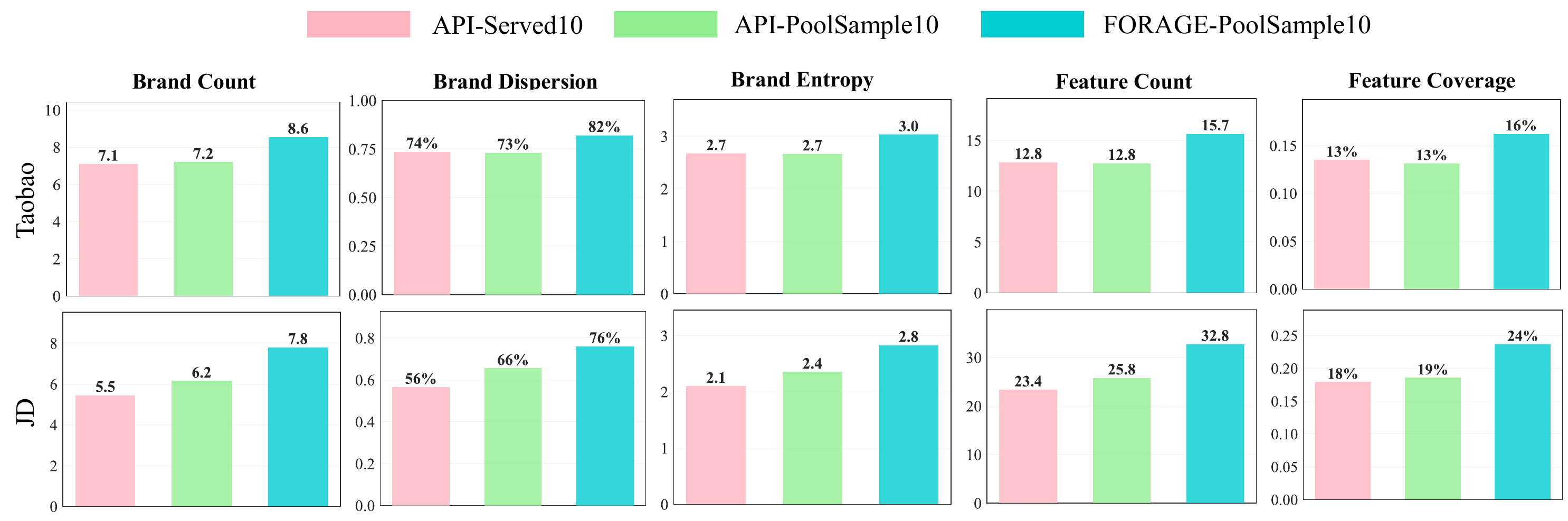}
    \caption{\textbf{Top-10 candidate-pool diversity.} \textsc{API-Served10} denotes the platform's Top-10 results, while \textsc{API-PoolSample10} and \textsc{Forage-PoolSample10} denote equally sized samples drawn from the API and \algdiv\ Top-50 candidate pools.}
    \label{fig:top10}
\end{figure*}

RQ2 examines \algdiv\ from two complementary perspectives. First, we evaluate whether it improves the intrinsic diversity of the retrieved candidate pool. Figure~\ref{fig:top10} compares the platform's directly served Top-10 results with equally sized samples drawn from the API Top-50 pool and the budget-matched Top-50 pool constructed by \algdiv. Across both datasets, samples from the \algdiv\ pool exhibit substantially higher brand and feature diversity than those from the API pool, showing that the improvement originates from retrieval rather than from a diversity-aware downstream selector.

Second, to verify that the gains originate from \algdiv's query-selection strategy rather than additional retrieval calls, we compare it with alternative probing methods under a matched multi-round retrieval protocol. As shown in Table~\ref{tab:keyword}, \algdiv\ outperforms the competing strategies in both retrieval diversity and recommendation quality. Incorporating an LLM actor into inverse-frequency probing improves performance but still falls short of \algdiv, highlighting the additional benefit of graph-guided exposure estimation beyond frequency-based selection and semantic filtering.

\begin{table}[t]
\centering
\small
\setlength{\tabcolsep}{4.5pt}
\renewcommand{\arraystretch}{1.1}
\caption{
Retrieval diversity and end-to-end recommendation performance under a matched
multi-round retrieval-call and candidate-pool budget, using the Qwen3.5-122B-A10B backbone with the Vanilla recommender on JD datasets. B-Ent denotes Brand Entropy, and F-Cov denotes Feature Coverage.
}
\label{tab:keyword}
\resizebox{\columnwidth}{!}{%
\begin{tabular}{l cccc}
\toprule
Query strategy
& B-Ent{$~\uparrow$}
& F-Cov{$~\uparrow$}
& Precision@10
& NDCG@10 \\
\midrule

Random Probe
& 2.24
& 16\%
& 0.5512
& 0.7203\\

Inverse-Freq (IF) Probe
& 2.46
& 19\%
& 0.5566
& 0.7286\\

LLM-Generated Probe
& 2.51
& 17\%
& 0.5530
& 0.7237\\

MMR-based Probe
& 2.44
& 17\%
& 0.5522
& 0.7233\\

IF Probe + LLM actor
& 2.72
& 21\%
& 0.5664
& 0.7352\\

\midrule

\textbf{\algdiv}
& \textbf{2.82}
& \textbf{24\%}
& \textbf{0.5713}
& \textbf{0.7426}\\
\bottomrule
\end{tabular}%
}
\end{table}

\subsection{Rerank Defense Effectiveness (RQ3)}
\label{sec:exp:rq3}

We evaluate the robustness of \algcom\ against review manipulation by comparing it against no defense, guarded prompting, and the GEO defenses SCI-Defense and SafeGEO. We consider two types of attacks: attribute-free subjective-text injection and fabricated attribute-claim injection. We measure the rank promotion ($\Delta$ Rank) of the lowest-ranked target item as the injection ratio increases.

\begin{figure}[t]
    \centering
    \includegraphics[width=\linewidth]{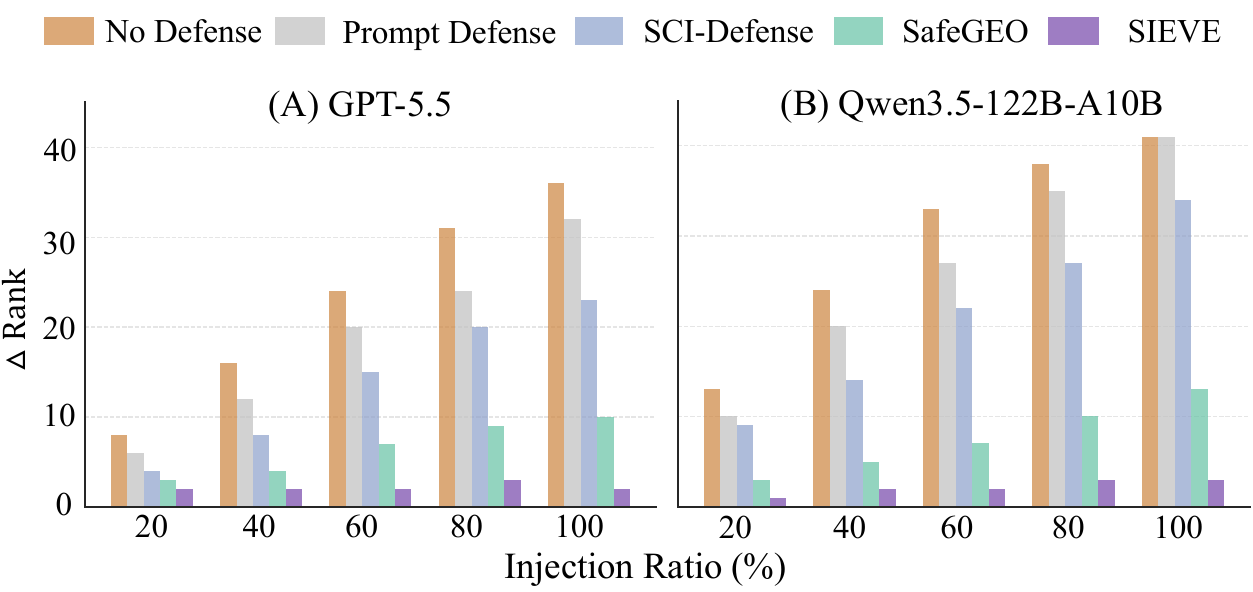}
    \caption{Rank promotion ($\Delta$ Rank) under attribute-free subjective-text injection with different defenses.}
    \label{fig:rq3}
\end{figure}

\noindent\textbf{Attribute-free subjective-text injection.}
Figure~\ref{fig:rq3} compares different defenses against subjective-text injection, where additional text is appended to existing reviews without introducing new attribute evidence. Without defense, increasing the injection ratio substantially promotes the target item. Guarded prompting provides partial mitigation but becomes less effective as subjective volume increases. The adapted GEO defenses also exhibit larger ranking shifts than \algcom. These results demonstrate the effectiveness of \algcom against attribute-free review manipulation. LLM-generated and defense-aware attacks are in Appendix~\ref{app:attack_variants}.

\noindent\textbf{Fabricated attribute-claim injection.}
We further evaluate a stronger attack that inserts LLM-generated, category-relevant attribute claims into existing reviews. As shown in Table~\ref{tab:fabricated_attack}, existing defenses provide limited protection, with ranking manipulation increasing substantially as more reviews are injected. In contrast, \algcom\ reduces the additional rank promotion caused by increasing injection volume, although fabricated claims can still influence the final ranking. This mitigation arises from set-based evidence consolidation, which prevents semantically overlapping claims from accumulating additional weight. Since \algcom\ does not verify factual correctness, these results demonstrate resistance to evidence-volume amplification rather than protection against fabricated claims.

\begin{table}[t]
    \centering
    \small
    \setlength{\tabcolsep}{3pt}
    \caption{Rank promotion ($\Delta$ Rank $\downarrow$) under fabricated attribute-claim injection at 20\%/60\%/100\% injection ratios.}
    \label{tab:fabricated_attack}
    \resizebox{\columnwidth}{!}{%
    \begin{tabular}{lccccc}
        \toprule
        Backbone & Vanilla & Prompt & SCI-Defense & SafeGEO & \textbf{\algcom} \\
        \midrule
        GPT-5.5
        & 13/28/40 & 12/25/37 & 13/23/32 & 10/21/30 & \textbf{10/13/14} \\
        Qwen3.5
        & 18/33/44 & 16/24/39 & 14/20/35 & 13/22/30 & \textbf{12/14/16} \\
        \bottomrule
    \end{tabular}%
    }
\end{table}

\subsection{Ablation Study (RQ4)}
\label{sec:exp:rq4}

\noindent\textbf{Module-level ablations.}
Table~\ref{tab:ablation} reports the ablation results on Taobao. Removing either \algdiv\ or \algcom\ reduces recommendation performance, with a larger degradation observed without \algdiv. This is because \algdiv\ expands candidate coverage, allowing the reranker to access more relevant products, while \algcom\ improves evidence quality by filtering attribute-free content and consolidating repeated claims. These results confirm the complementary contributions of retrieval diversification and evidence consolidation.

\begin{table}[t]
\centering
\caption{Module-level ablation of \method\ on Taobao using both backbones (RQ4). The best results are in \textbf{bold}.}
\label{tab:ablation}
\small
\setlength{\tabcolsep}{3.5pt}
\renewcommand{\arraystretch}{1.05}
\resizebox{\columnwidth}{!}{%
\begin{tabular}{@{}lcccc@{}}
\toprule
& \multicolumn{2}{c}{GPT-5.5}
& \multicolumn{2}{c}{Qwen3.5-122B-A10B} \\
\cmidrule(lr){2-3}
\cmidrule(lr){4-5}
Method
& Precision@10
& NDCG@10
& Precision@10
& NDCG@10 \\
\midrule
\method
& \textbf{0.5822}
& \textbf{0.7465}
& \textbf{0.5513}
& \textbf{0.7233} \\
\quad w/o \algcom
& 0.5746
& 0.7403
& 0.5427
& 0.7142 \\
\quad w/o \algdiv
& 0.5714
& 0.7260
& 0.5368
& 0.7056 \\
Vanilla
& 0.5597
& 0.7236
& 0.5334
& 0.7019 \\
\bottomrule
\end{tabular}%
}
\vspace{-2mm}
\end{table}

\noindent\textbf{\algdiv\ internal ablations.}
We further isolate the roles of the graph advisor and the LLM actor. As shown in Table~\ref{tab:forage_ablation}, removing either component reduces retrieval diversity and downstream recommendation performance. The graph advisor identifies structurally underrepresented features, while the LLM actor selects executable query anchors from the advised candidates. These results support their complementary roles in \algdiv.

\begin{table}[t]
\centering
\small
\setlength{\tabcolsep}{4.5pt}
\renewcommand{\arraystretch}{1.1}
\caption{Internal ablation of \algdiv\ on JD using the Vanilla recommender and Qwen3.5-122B-A10B.}
\label{tab:forage_ablation}
\resizebox{\columnwidth}{!}{%
\begin{tabular}{lcccc}
\toprule
Query strategy
& B-Ent{$~\uparrow$}
& F-Cov{$~\uparrow$}
& Precision@10
& NDCG@10 \\
\midrule
\algdiv\ (w/o LLM Actor)
& 2.62 & 21\% & 0.5663 & 0.7355 \\
\algdiv\ (w/o Graph Advisor)
& 2.69 & 21\% & 0.5673 & 0.7361 \\
\textbf{\algdiv}
& \textbf{2.82} & \textbf{24\%} & \textbf{0.5713} & \textbf{0.7426} \\
\bottomrule
\end{tabular}%
}
\vspace{-2mm}
\end{table}

\noindent\textbf{\algcom\ internal ablation.}
We also remove Set-based Evidence Consolidation (SetEviCon) while retaining claim extraction. As shown in Table~\ref{tab:sieve_ablation}, removing SetEviCon reduces recommendation quality and substantially increases attack-induced ranking shifts on both backbones. These results demonstrate the importance of set-based consolidation for both recommendation quality and robustness.

\begin{table}[t]
    \centering
    \small
    \setlength{\tabcolsep}{4pt}
    \caption{Internal ablation of \algcom\ within the full \method\ pipeline on Taobao. Rank promotion ($\Delta$ Rank $\downarrow$) is measured under attribute-free injection at 100\%.}
    \label{tab:sieve_ablation}
    \resizebox{\columnwidth}{!}{%
    \begin{tabular}{lcccccc}
        \toprule
        & \multicolumn{3}{c}{GPT-5.5}
        & \multicolumn{3}{c}{Qwen3.5-122B-A10B} \\
        \cmidrule(lr){2-4} \cmidrule(lr){5-7}
        Method & Precision@10 & NDCG@10 & $\Delta$ Rank $\downarrow$
        & Precision@10 & NDCG@10 & $\Delta$ Rank $\downarrow$ \\
        \midrule
        \method
        & \textbf{0.5822} & \textbf{0.7465} & \textbf{2}
        & \textbf{0.5513} & \textbf{0.7233} & \textbf{2} \\
        \quad w/o SetEviCon
        & 0.5773 & 0.7430 & 13
        & 0.5469 & 0.7186 & 17 \\
        \bottomrule
    \end{tabular}%
    }
\vspace{-2mm}
\end{table}

\subsection{Efficiency Analysis (RQ5)}
\label{sec:exp:rq5}

Table~\ref{tab:eff} reports the effectiveness--efficiency trade-off of \method. We measure the complete recommendation trajectory, counting both prompt and completion tokens and all retrieved metadata, reviews, and intermediate evidence supplied to the models. The additional cost reflects the computation required for retrieval diversification and evidence consolidation, rather than longer final responses or redundant generation. To further examine the quality--cost trade-off, we vary the number of retrieval rounds in the full \method\ pipeline. As the search budget increases, recommendation quality consistently improves, accompanied by higher token consumption and latency. These results suggest that additional retrieval rounds provide meaningful gains, while reduced-budget configurations offer more efficient alternatives when computational resources are limited. In addition, \algcom\ provides robustness against evidence manipulation, a benefit not fully captured by clean recommendation metrics. Since its evidence profiles can be precomputed and reused, the online cost may be further reduced in deployment.

\begin{table}[t]
    \centering
    \small
    \setlength{\tabcolsep}{3.5pt}
    \caption{Effectiveness--efficiency trade-off with GPT-5.5 on Taobao.
    Total tokens are accumulated over LLM calls.}
    \label{tab:eff}
    \resizebox{\columnwidth}{!}{%
    \begin{tabular}{lcccccc}
        \toprule
        Method
        & \shortstack{\# Search calls}
        & \shortstack{\# LLM calls}
        & Tokens
        & Latency (s)
        & Precision@10
        & NDCG@10 \\
        \midrule
        Base
        & $1$ & $1$ & $\sim 87$K
        & $20$ & $0.5597$ & $0.7236$ \\
        \quad +\,\algdiv
        & $5$ & $5$ & $\sim 134$K
        & $32$ & $0.5746$ & $0.7403$ \\
        \quad +\,\algcom
        & $1$ & $2$ & $\sim 150$K
        & $23$ & $0.5714$ & $0.7260$ \\
        \midrule
        \method (\# S = 2)
        & $2$ & $3$ & $\sim 162$K
        & $26$ & $0.5749$
        & $0.7301$ \\
        \method (\# S = 3)
        & $3$ & $4$ & $\sim 174$K
        & $29$ & $0.5776$
        & $0.7357$ \\
        \method (\# S = 4)
        & $4$ & $5$ & $\sim 185$K
        & $31$ & $0.5803$
        & $0.7421$ \\
        \method
        & $5$ & $6$ & $\sim 197$K
        & $34$ & $\mathbf{0.5822}$
        & $\mathbf{0.7465}$ \\
        \bottomrule
    \end{tabular}%
    }
    \vspace{-4mm}
\end{table}

\section{CONCLUSION}

We present \method, a defense harness for agentic recommender systems that addresses two distinct failure modes: retrieval-stage diversity deficits that constrain candidate exposure and reranking-stage review manipulation that can distort product rankings. We address these issues with \algdiv\ for graph-guided retrieval diversification and \algcom\ for attribute-grounded evidence consolidation. Together, these modules broaden candidate exposure and reduce the amplification of manipulated review content, improving recommendation quality while mitigating ranking manipulation under the evaluated review-text injection attacks.

\bibliographystyle{ACM-Reference-Format}
\bibliography{ref}

\appendix
\section{APPENDIX}

\subsection{Benchmark Construction}
\label{app:data-construction}

\subsubsection{Benchmark description.}
We construct the benchmark \bench from two e-commerce platforms, JD and Taobao. The same 20 category-level queries are issued on both platforms. For category $c$ and platform $s\in \{\text{JD},\text{Taobao}\}$, one benchmark instance is represented as
\begin{equation}
  x_{c,s}=\left(q_c,\pi_{c,s}\right), \qquad
  \pi_{c,s}=\left(p_{c,s,1},\ldots,p_{c,s,K_{c,s}}\right),
  \label{eq:benchmark-instance}
\end{equation}
where $q_c$ is the category query, $\pi_{c,s}$ is the platform-native ordered product sequence, $p_{c,s,r}$ is the product at rank $r$, and $K_{c,s}$ is the number of products stored for that platform-query pair. Each product is linked to its price, available structured metadata, and review text collected from its product page. 

\subsubsection{Agent-based data acquisition.} Data acquisition is conducted by \textsc{GenericAgent}~\cite{liang2026genericagent} under a researcher-defined standard operating procedure (SOP). The SOP specifies the target product categories, the fields of research interest, the navigation sequence, and the stopping conditions. All operational steps of the acquisition process are performed by the agent. Specifically, \textsc{GenericAgent} locates the requested category, opens the corresponding product pages, identifies the product and review sections, and records the platform responses associated with the predefined fields. Because additional reviews are rendered only after the review panel is scrolled, the agent autonomously scrolls the panel, detects newly loaded entries, and continues the extraction process until the predefined collection condition is satisfied. The agent interacts only with product and review pages available through the platform's standard user interface and does not access private messages, purchase histories, payment information, or other private account areas.

\begin{table}[t]
  \caption{Product categories included in the benchmark.}
  \label{tab:product-categories}
  \centering
  \footnotesize
  \setlength{\tabcolsep}{4pt}
  \begin{tabular}{@{}p{0.46\columnwidth}p{0.46\columnwidth}@{}}
    \toprule
    1. Ergonomic Chair  & 11. Blender \\
    2. Tablet           & 12. Air Fryer \\
    3. Robot Vacuum     & 13. Air Conditioner \\
    4. Tissue           & 14. Laptop \\
    5. Monitor          & 15. Massage Gun \\
    6. Pillow           & 16. Bluetooth Earbuds \\
    7. Shampoo          & 17. Sneakers \\
    8. Washing Machine  & 18. Keyboard \\
    9. Electric Kettle  & 19. Sunscreen \\
    10. Camera          & 20. Dehumidifier \\
    \bottomrule
  \end{tabular}
\end{table}

\subsubsection{Dataset composition and evidence coverage.}
The benchmark contains 40 query-conditioned product lists collected from
JD.com (hereafter JD) and Taobao across 20 aligned product categories. These
lists comprise 2,086 ranked query--product entries and 2,085 platform-specific
products. Each list contains between 46 and 60 products, with an average length
of 52.15. Products are identified independently within each platform, and no
cross-platform entity resolution is performed. The benchmark further provides
131,459 product-linked review records. JD contributes 1,140 products and
46,549 review records, while Taobao contributes 945 products and 84,910 review
records. At least one review record is available for 1,923 products,
corresponding to an overall product-level review coverage of 92.23\%. Across
the complete benchmark, each product is associated with an average of 63.05
review records and 14.12 non-empty structured metadata fields.
Tables~\ref{tab:corpus-scale} and~\ref{tab:corpus-density} summarize the
platform-level scale and evidence density.

\begin{table}[t]
  \caption{Platform-level scale of the benchmark. Products are counted
  independently within JD and Taobao.}
  \label{tab:corpus-scale}
  \centering
  \small
  \setlength{\tabcolsep}{6pt}
  \begin{tabular}{@{}lrr@{}}
    \toprule
    Source & \#Products & \#Review Records \\
    \midrule
    JD      & 1,140 & 46,549 \\
    Taobao  &   945 & 84,910 \\
    \bottomrule
  \end{tabular}
\end{table}

\begin{table}[H]
  \caption{Average product-level evidence density. Products without collected
  review records are included with zero reviews when computing the averages.}
  \label{tab:corpus-density}
  \centering
  \small
  \setlength{\tabcolsep}{4pt}
  \begin{tabular}{@{}lrr@{}}
    \toprule
    Source
      & \shortstack{Avg. review records\\per product}
      & \shortstack{Avg. metadata fields\\per product} \\
    \midrule
    JD      & 40.83 &  7.94 \\
    Taobao  & 89.85 & 21.58 \\
    \bottomrule
  \end{tabular}
\end{table}

Because JD and Taobao expose different product-attribute schemas, the metadata
counts characterize the amount of structured information available within
each platform rather than directly comparable semantic feature dimensions.
Moreover, the updated archive does not retain stable review identifiers.
Therefore, the reported review totals refer to stored review records and
should not be interpreted as the number of unique reviewers.

\subsubsection{Platform-Provided Brand and Attribute Catalogs}
\label{app:catalog}

For each platform $s$ and product category $c$, the collected dataset contains the complete category-level brand catalog $\mathcal{B}_{s,c}$, the complete category-level attribute vocabulary $\mathcal{V}_{s,c}$, and the structured brand and attribute metadata attached by the platform to each product.

For a product $i$ belonging to category $c$ on platform $s$, its brand $b(i)\in\mathcal{B}_{s,c}$ and attribute set $\phi(i)\subseteq\mathcal{V}_{s,c}$ are read directly from the structured fields exposed by the platform, without extracting or inferring additional attributes from its title, description, reviews, images, or LLM-generated annotations.

We define the platform-exposed reference catalog for each platform-category pair as $\mathcal{M}_{s,c}=(\mathcal{B}_{s,c},\mathcal{V}_{s,c})$, where ``complete'' refers to the full brand and attribute catalogs disclosed by the platform for that category at the time of data collection rather than the entire real-world market.

The reference catalogs are not constructed from the union of initially retrieved products, adaptively retrieved products, or products returned by any particular retrieval strategy, and therefore their contents and sizes remain fixed across all evaluated methods.

During retrieval, neither \algdiv nor any baseline has access to the global catalogs $\mathcal{B}_{s,c}$ or $\mathcal{V}_{s,c}$, and each policy observes only the brand, structured attributes, and reviews associated with products already returned by the black-box retrieval tool.

Accordingly, at retrieval round $t$, the locally observable brand and attribute spaces are defined as
$\mathcal{B}_t=\{b(i)\mid i\in\mathcal{P}_t\}$ and
$\mathcal{V}_t=\bigcup_{i\in\mathcal{P}_t}\phi(i)$,
both of which may cover only a subset of the corresponding platform-provided reference catalogs.

In particular, the graph advisor constructs its item--feature incidence matrix exclusively from $\mathcal{P}_t$ and $\mathcal{V}_t$, while the LLM actor is restricted to the graph-advised candidates drawn from $\mathcal{V}_t$, preventing either component from querying attributes that are revealed only by the global reference vocabulary.

For offline evaluation, the feature coverage of a candidate pool $\mathcal{P}$ is computed as
$\operatorname{cov}(\mathcal{P})=
\left|\bigcup_{i\in\mathcal{P}}\phi(i)\right|/
|\mathcal{V}_{s,c}|$,
so that all retrieval strategies are evaluated against the same platform-provided denominator.

Brand-based metrics are computed from the platform-provided product brands observed in the evaluated pool, while $\mathcal{B}_{s,c}$ provides the fixed category-level reference used to characterize how much of the disclosed brand space is represented.

Because the catalogs are used only after retrieval for metric computation, they do not provide additional search terms, candidate identities, or other privileged information to \algdiv, its ablations, or the competing probing strategies.

For readability, the main text omits the platform and category subscripts and writes $\mathcal{M}=(\mathcal{B},\mathcal{V})$, with the corresponding platform-category pair understood from the evaluation instance.

\subsubsection{Capability comparison with related datasets.}
Table~\ref{tab:dataset-capability-comparison} compares the extent to
which each released resource directly supports the six data properties
required by our benchmark. A checkmark indicates that the corresponding
property is explicitly available in the released data, while a cross
indicates that external data, heuristic inference, or additional
reconstruction would be required.

A query-conditioned product list associates candidate products with a
specific search query. Native order refers to the original order in which
the products were displayed or exposed by the platform. Cross-platform
pairing requires aligned product-category queries to be executed on at
least two independent platforms. Structured attributes refer to
candidate-level product fields beyond identifiers and titles, while
candidate-level review text requires review content to be explicitly
linked to each returned product. An explicit ad indicator requires a
candidate-level field that directly identifies a returned product as an
advertising or sponsored result. Query-source fields, promotion-related
text, relevance labels, and behavioral outcomes such as clicks or
purchases are not treated as ad indicators.

The comparison focuses on direct task support rather than raw dataset
scale. Although several existing resources contain substantially more
queries, products, or reviews, none of the compared datasets jointly
provides platform-native product lists, cross-platform query alignment,
structured product attributes, candidate-linked reviews, and explicit
advertising labels.

Specifically, ESCI~\cite{reddy2022shopping} and
WANDS~\cite{chen2022wands} support query--product relevance assessment,
but neither dataset releases platform-native rankings, candidate-level
review text, or explicit advertising labels. JDsearch~\cite{liu2023jdsearch}
and KuaiSearch~\cite{li2026kuaisearch} preserve real platform search
interactions and product sequences. However, their released schemas
provide neither reviews linked to individual candidates nor fields that
distinguish sponsored products from organic results. Their interaction
labels describe user behaviors, such as clicks, add-to-cart actions, and
purchases, rather than advertising status.

TREC Product Search 2023~\cite{campos2023trecproduct} provides queries,
structured product information, and review-related evidence, but its
ranked lists are generated by retrieval systems rather than obtained as
aligned native rankings from multiple commercial platforms. It also does
not expose an explicit sponsored-result indicator. Amazon Reviews
2023~\cite{hou2024amazonreviews} provides product metadata and review text
at substantially larger scale, but it contains neither query-conditioned
product sequences nor advertising labels.

In contrast, our benchmark directly retains explicit candidate-level
advertising indicators from the platform responses: \texttt{isAdv} for
JD and \texttt{isAd} for Taobao. This information supports separate
analysis of sponsored and organic exposure, which cannot be performed
directly with the compared datasets without collecting additional
platform data or inferring advertising status through potentially
unreliable heuristics. Consequently, among the compared released
resources, our benchmark is the only one that directly provides all six
properties.

\providecommand{\cmark}{\ding{51}}
\providecommand{\xmark}{\ding{55}}

\begin{table*}[t]
\caption{Capability comparison with representative product-search and review datasets. Native order indicates that the released resource preserves the platform's displayed or exposed product order. An explicit ad indicator requires a released candidate-level field that directly identifies an advertising or sponsored result.}
\label{tab:dataset-capability-comparison}

\centering
\small
\setlength{\tabcolsep}{4pt}
\renewcommand{\arraystretch}{1.18}

\begin{tabularx}{\textwidth}{
  @{}
  >{\raggedright\arraybackslash}p{0.245\textwidth}
  *{6}{>{\centering\arraybackslash}X}
  @{}
}
\toprule
\textbf{Dataset}
  & \shortstack{\textbf{Query-conditioned}\\\textbf{product lists}}
  & \shortstack{\textbf{Released}\\\textbf{native order}}
  & \shortstack{\textbf{Cross-platform}\\\textbf{paired queries}}
  & \shortstack{\textbf{Structured}\\\textbf{attributes}}
  & \shortstack{\textbf{Candidate-level}\\\textbf{review text}}
  & \shortstack{\textbf{Explicit}\\\textbf{ad indicator}} \\
\midrule

ESCI~\cite{reddy2022shopping}
  & \cmark & \xmark & \xmark & \cmark & \xmark & \xmark \\

WANDS~\cite{chen2022wands}
  & \cmark & \xmark & \xmark & \cmark & \xmark & \xmark \\

JDsearch~\cite{liu2023jdsearch}
  & \cmark & \cmark & \xmark & \cmark & \xmark & \xmark \\

KuaiSearch~\cite{li2026kuaisearch}
  & \cmark & \cmark & \xmark & \cmark & \xmark & \xmark \\

TREC Product Search 2023~\cite{campos2023trecproduct}
  & \cmark & \xmark & \xmark & \cmark & \cmark & \xmark \\

Amazon Reviews 2023~\cite{hou2024amazonreviews}
  & \xmark & \xmark & \xmark & \cmark & \cmark & \xmark \\

\midrule
\bench
  & \cmark & \cmark & \cmark & \cmark & \cmark & \cmark \\

\bottomrule
\end{tabularx}
\end{table*}

\subsubsection{Privacy-preserving data release.}
The publicly released benchmark is a privacy-filtered version of the raw platform data rather than a direct copy. Although the collection targets product information and review text, the raw records may also contain reviewer nicknames, review identifiers, exact timestamps, account-level indicators, and personal information disclosed in free-form reviews. Before release, we remove all unnecessary reviewer-side and account-linking fields, screen the review text for potentially sensitive information, and manually inspect flagged records. Confirmed personal identifiers are redacted, and records that cannot be safely processed are excluded. The raw archive remains restricted to the research team, while the released benchmark contains only the information required for the evaluation tasks. The code and dataset will be released soon.

\subsubsection{Ethical statement.} 
This study does not recruit, contact, or interact with individual reviewers. Reviewer-level attributes are neither used as model inputs nor training targets, and are not considered in evaluation. Reviews are accessed through a valid, authenticated browser session, while page navigation, scrolling, and data extraction are performed by the agent. Authenticated access is treated solely as a collection mechanism and does not constitute reviewer consent or permission to redistribute user-generated content. Accordingly, unnecessary reviewer-related fields are removed before release, and the benchmark is limited to aggregate research on query-conditioned product lists. It is not intended for reviewer identification, behavioral profiling, surveillance, or commercial targeting. Data collection, storage, and release are conducted in accordance with applicable platform policies and institutional ethics requirements.

\subsection{Evaluation Metrics}
\label{app:evaluation-metrics}
We evaluate each Top-$K$ recommendation list from two complementary perspectives. \emph{Recommendation quality} measures whether the returned products match the user query and persona, whereas \emph{retrieval diversity} measures the breadth and balance of the brands and product features represented in the retrieved set. Recommendation-quality metrics are computed for each user-query instance and averaged over all evaluated instances, while retrieval-diversity metrics are computed for each product category and macro-averaged across the 20 categories.

\subsubsection{Recommendation Quality}
\label{app:recommendation-quality}

\begin{figure}[t]
  \centering
  \includegraphics[width=\linewidth]{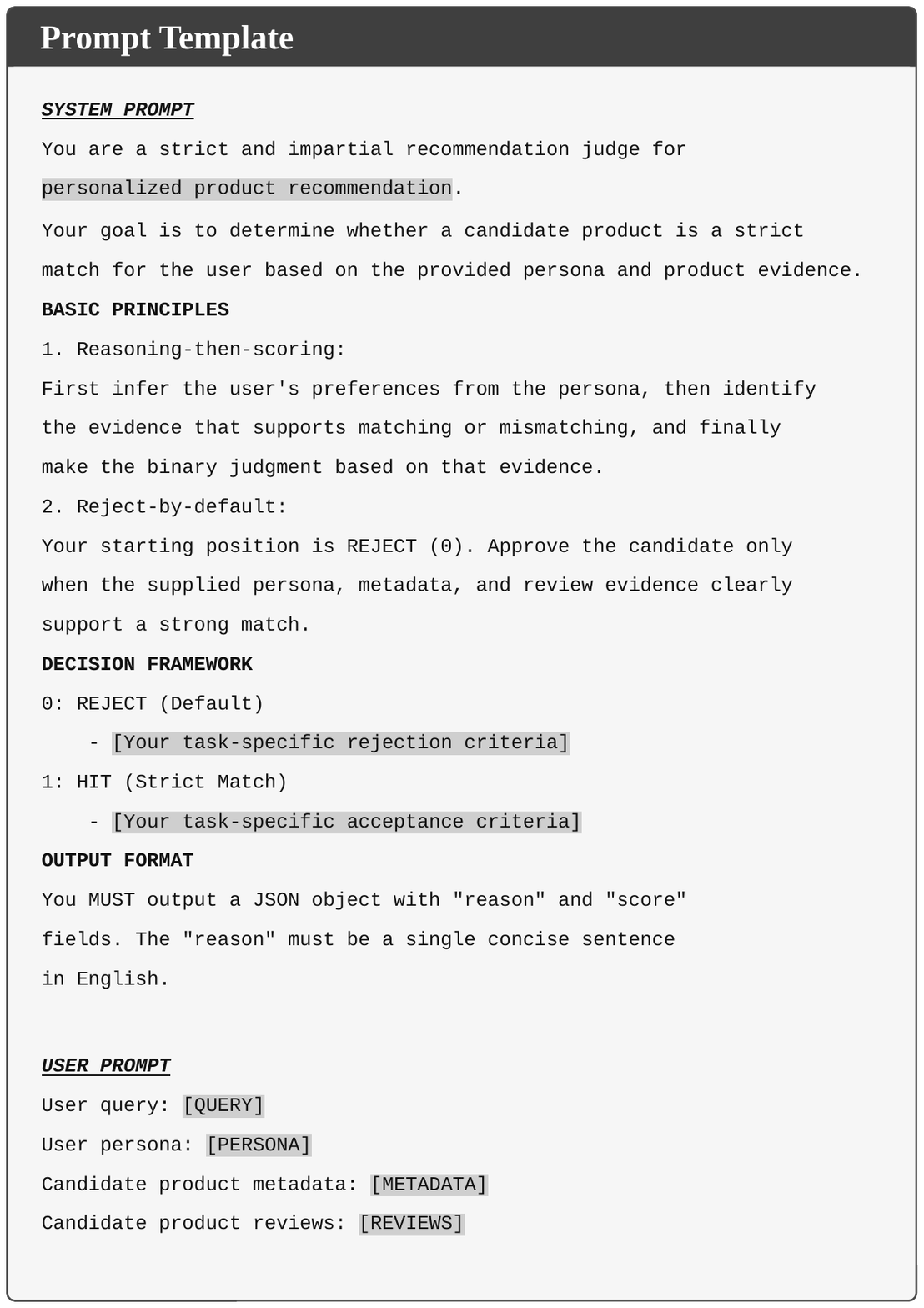}
  \caption{Evaluation prompt applied in our LLM Judge. The basic principles and output format remain fixed, while the input fields are instantiated for each query--persona--product evaluation instance.}
  \label{fig:product-judge-prompt}
  \Description{The evaluation prompt used by the LLM Judge. The basic principles and output format are fixed, while the shaded fields contain the query, persona, product metadata, and a method-independent raw-review evidence packet.}
\end{figure}

Following the pointwise LLM-as-a-Judge framework~\cite{que2026llm}, the judge evaluates each recommended product independently using the user query, user persona, product metadata, and a fixed raw-review evidence packet. The judge is based on DeepSeek-V4-Pro. The prompt follows a reasoning-then-scoring procedure and a reject-by-default policy. The judge first infers the user's preferences from the persona, identifies evidence supporting a match or mismatch from the supplied product information, and then returns a JSON object containing a one-sentence rationale and a binary relevance score. A score of $1$ denotes a strict, evidence-supported match, whereas $0$ denotes an irrelevant, conflicting, or insufficiently supported recommendation. The judge receives no method-specific information, including the recommender system, backbone, ranking process, or intermediate evidence representation. Only the binary score is used for quantitative evaluation. Figure~\ref{fig:product-judge-prompt} shows the prompt template used in our product-recommendation setting.

Formally, let $\mathcal{Q}$ denote the set of evaluated query--persona instances and let $\mathcal{M}$ denote the set of evaluated recommendation methods. For an instance $q\in\mathcal{Q}$, let $p_q$ denote its persona. For method $m\in\mathcal{M}$, let $\pi_K^{(m)}(q)=(i_{q,1}^{(m)},\ldots,i_{q,K}^{(m)})$ denote its ordered Top-$K$ recommendation list. For each product $i$, let $M_i$ denote its structured metadata and let $\widetilde{\mathcal{R}}_i$ denote its fixed raw-review evidence packet. The judge assigns the binary relevance label
\[
  y_{q,r}^{(m)}
  =
  \operatorname{Judge}
  \left(
    q,
    p_q,
    M_{i_{q,r}^{(m)}},
    \widetilde{\mathcal{R}}_{i_{q,r}^{(m)}}
  \right)
  \in\{0,1\}.
\]
Because neither the recommendation method nor a method-specific evidence representation appears in the judge input, the same relevance label is reused whenever the same query--persona--product tuple occurs in multiple recommendation lists.

Precision@$K$ measures the proportion of judged-relevant products in the returned list:
\begin{equation}
  \mathrm{Precision}^{(m)}@K
  =
  \frac{1}{|\mathcal{Q}|}
  \sum_{q\in\mathcal{Q}}
  \frac{1}{K}
  \sum_{r=1}^{K}
  y_{q,r}^{(m)}.
  \label{eq:judge-precision}
\end{equation}

To account for ranking position, we use the same binary relevance labels in discounted cumulative gain and construct a method-independent evaluation universe from the union of products returned by all compared methods.

For each query--persona instance $q$, let $\mathcal{U}_q=\bigcup_{m\in\mathcal{M}}\{i_{q,1}^{(m)},\ldots,i_{q,K}^{(m)}\}$ denote the set of unique products appearing in at least one evaluated Top-$K$ recommendation list.

Each query--persona--product tuple $(q,p_q,i)$ with $i\in\mathcal{U}_q$ is judged once using the same structured metadata and fixed raw-review evidence packet, yielding the method-independent binary relevance label $y_{q,i}=\operatorname{Judge}(q,p_q,M_i,\widetilde{\mathcal{R}}_i)\in\{0,1\}$.

For a product appearing at rank $r$ in the output of method $m$, we define $y_{q,r}^{(m)}=y_{q,i_{q,r}^{(m)}}$, so the same relevance label is reused whenever the same product appears in multiple recommendation lists.

We compute the discounted cumulative gain of method $m$ as
\begin{equation}
  \mathrm{DCG}^{(m)}@K(q)=\sum_{r=1}^{K}\frac{2^{y_{q,r}^{(m)}}-1}{\log_2(r+1)}.
  \label{eq:judge-dcg}
\end{equation}

Let $H_q=\min\left(K,\sum_{i\in\mathcal{U}_q}y_{q,i}\right)$ denote the maximum number of judged-relevant products that can be placed within the first $K$ positions under the shared evaluation universe.

We define the method-independent ideal discounted cumulative gain as
\begin{equation}
  \mathrm{IDCG}@K(q)=\sum_{r=1}^{H_q}\frac{1}{\log_2(r+1)}.
  \label{eq:judge-idcg}
\end{equation}

The normalized discounted cumulative gain of method $m$ is computed as
\begin{equation}
  \mathrm{NDCG}^{(m)}@K=\frac{1}{|\mathcal{Q}|}\sum_{q\in\mathcal{Q}}\frac{\mathrm{DCG}^{(m)}@K(q)}{\mathrm{IDCG}@K(q)},
  \label{eq:judge-ndcg}
\end{equation}
where the ratio is defined as zero when $H_q=0$.

\subsubsection{Retrieval Diversity}
\label{app:retrieval-diversity}
Retrieval diversity measures how broadly a retrieved product set covers brands and product features. We evaluate it through \emph{brand diversity} and \emph{feature diversity}. All five metrics are oriented so that larger values indicate greater diversity.

\paragraph{Compared retrieval sets.}
For category $c$, let $P_c^{\mathrm{API}}$ denote the ordered Top-50 pool returned by the platform API and $P_c^{\mathrm{MR}}$ the Top-50 pool obtained by multi-round retrieval. For an ordered pool $P$, $P[1{:}K]$ denotes its first $K$ products. With $K=10$, the compared sets are
\begin{align}
  S_c^{\mathrm{API}}
    &= P_c^{\mathrm{API}}[1{:}K],
    \label{eq:api-topk-set}\\
  S_c^{\mathrm{RR}}
    &\sim \operatorname{SampleWOR}(P_c^{\mathrm{API}},K),
    \label{eq:rr-topk-set}\\
  S_c^{\mathrm{OURS}}
    &\sim \operatorname{SampleWOR}(P_c^{\mathrm{MR}},K),
    \label{eq:ours-topk-set}
\end{align}
where $\operatorname{SampleWOR}(P,K)$ denotes uniform sampling of $K$ products from $P$ without replacement. These sets correspond to API-Served10, API-PoolSample10, and \algdiv-PoolSample10, respectively. To reduce the variance introduced by random subset selection, we independently sample $K$ products without replacement from each Top-50 pool over $B=1000$ repetitions. The same metrics are also applied directly to $P_c^{\mathrm{API}}$ and $P_c^{\mathrm{MR}}$ in the Top-50 comparison. In the following definitions, $S_c$ denotes any evaluated product set for category $c$, and $N_c=|S_c|$ is its size.

\paragraph{Brand diversity.}
Let $\beta(i)$ denote the brand of product $i$, and define the represented brand set, brand frequency, brand share, and three diversity metrics as
\begin{align}
  \mathcal{B}_c(S_c)
  &= \{\beta(i)\mid i\in S_c\},
  \label{eq:represented-brands}\\
  n_{c,b}(S_c)
  &= \sum_{i\in S_c}\mathbf{1}[\beta(i)=b],
  \label{eq:brand-frequency}\\
  p_{c,b}(S_c)
  &= \frac{n_{c,b}(S_c)}{N_c},\\
  \mathrm{BrandCount}_c(S_c)
  &= \left|\mathcal{B}_c(S_c)\right|,
  \label{eq:brand-count}\\
  \mathrm{BrandDispersion}_c(S_c)
  &= 1-\max_{b\in\mathcal{B}_c(S_c)}p_{c,b}(S_c),
  \label{eq:brand-dispersion}\\
  \mathrm{BrandEntropy}_c(S_c)
  &= -\sum_{b\in\mathcal{B}_c(S_c)}
    p_{c,b}(S_c)\log_2 p_{c,b}(S_c).
  \label{eq:brand-entropy}
\end{align}
Here, $n_{c,b}(S_c)$ is the number of products from brand $b$, and $p_{c,b}(S_c)$ is its share in $S_c$. \textbf{Brand Count} measures the number of distinct brands. \textbf{Brand Dispersion} equals zero when all products share one brand and reaches $1-1/N_c$ when every product has a different brand. \textbf{Brand Entropy}, measured in bits, captures both richness and exposure balance, ranging from zero for a single brand to $\log_2 N_c$ when every product has a different brand. Larger values indicate greater brand diversity for all three metrics.

\paragraph{Feature diversity.}
Let $\phi(i)$ denote the feature dimensions associated with product $i$ and $\mathcal{F}_c^{\mathrm{ref}}$ the complete reference feature set for category $c$. The represented feature set and two feature-diversity metrics are
\begin{align}
  \mathcal{F}_c(S_c)
  &= \bigcup_{i\in S_c}\phi(i),
  \label{eq:represented-features}\\
  \mathrm{FeatureCount}_c(S_c)
  &= \left|\mathcal{F}_c(S_c)\right|,
  \label{eq:feature-count}\\
  \mathrm{FeatureCoverage}_c(S_c)
  &= \frac{\left|\mathcal{F}_c(S_c)\right|}
          {\left|\mathcal{F}_c^{\mathrm{ref}}\right|}.
  \label{eq:feature-coverage}
\end{align}
\textbf{Feature Count} measures the absolute number of distinct feature dimensions covered by at least one retrieved product, whereas \textbf{Feature Coverage} normalizes this count by the category-level reference feature space and lies in $[0,1]$. A value of 1 indicates complete coverage. Unlike Feature Count, Feature Coverage is comparable across categories with different feature-space sizes.

\paragraph{Category-level aggregation.}
Each metric is computed independently for every category. Let $m_c$ be the value of any one of the five metrics for category $c$ and $C=20$ the number of categories. We report the macro-average
\begin{equation}
  \overline{m}
  = \frac{1}{C}\sum_{c=1}^{C}m_c
  = \frac{1}{20}\sum_{c=1}^{20}m_c.
  \label{eq:diversity-macro-average}
\end{equation}
This aggregation assigns equal weight to each category. Brand Count, Brand Entropy, and Feature Count are reported as raw macro-averaged values, while Brand Dispersion and Feature Coverage are reported as percentages.

\subsection{Brand Diversity Baselines for Observation~2}
\label{app:brand-diversity}

This section details how we compute the brand-concentration results in Observation~2 by comparing the brands exposed by the retrieval platform against both an equal-exposure baseline and a conventional long-tail market baseline.

\noindent\textbf{Metric.}
For an item set $\mathcal{S}$, we define its head-brand share as
\begin{equation}
H(\mathcal{S})
=
\max_{b}
\frac{\left|\{i\in\mathcal{S}:b(i)=b\}\right|}{|\mathcal{S}|},
\end{equation}
where $b(i)$ is the platform-provided brand label and a larger value indicates that a greater fraction of the visible items is occupied by one brand.

\noindent\textbf{Observed retrieval results.}
For each of the 20 product categories, we retain the first 50 items returned by the platform, estimate the retrieval-pool statistic by averaging $H(\mathcal{S})$ over 2,000 uniformly sampled subsets of ten items without replacement, and compute the API-served statistic directly from the first ten items in the platform-provided order.

\noindent\textbf{Uniform baseline.}
Given the $M$ brands exposed in the platform's full category catalog, we simulate 5,000 lists of $K=10$ items from a uniform categorical distribution and average their head-brand shares.

\noindent\textbf{Long-tail baseline.}
We construct the Zipf baseline over the same $M$-brand universe using
\begin{equation}
p_r
=
\frac{r^{-\hat{s}}}
{\sum_{j=1}^{M}j^{-\hat{s}}},
\end{equation}
where $\hat{s}$ is estimated from an independent market-side brand distribution rather than from the evaluated retrieval pool, and we then simulate 5,000 lists of ten items to obtain the expected head-brand share under natural long-tail concentration.

\noindent\textbf{Interpretation.}
We report the macro-averaged values
\begin{equation}
H_{\mathrm{Uniform}},
\quad
H_{\mathrm{Zipf}},
\quad
H_{\mathrm{Retrieval}},
\quad
H_{\mathrm{Top10}},
\end{equation}
where $H_{\mathrm{Retrieval}}>H_{\mathrm{Zipf}}$ indicates that the retrieved candidate pool is more brand-concentrated than expected from conventional market skew, while $H_{\mathrm{Top10}}>H_{\mathrm{Retrieval}}$ indicates that the platform ordering further amplifies this concentration.

\subsection{Measurement of Attribute-Grounded Review Content for Observation 3}
\label{app:review-content-analysis}

To quantify the proportion of review text that contains product-related attribute information, we formulate the analysis as an LLM-based span identification task. For each of the four representative categories shown in Figure~\ref{fig:evi_poi}A, we sample 50 products and retain 50 reviews for each product, resulting in 2,500 reviews per category. All reviews are processed using Deepseek-v4-pro~\cite{xu2026deepseek} under the same prompt and decoding settings.

Given an original review, the LLM segments the text into sentences or clauses and identifies the spans that explicitly describe a concrete product, fulfillment, or service attribute. A span is considered attribute-related when it connects an identifiable aspect, such as material, durability, comfort, operational noise, delivery, or after-sales service, with a factual description or evaluative statement. Sentences containing only generic praise, emotional endorsement, recommendation, popularity, urgency, or repurchase intent are not selected. The LLM is required to return the selected spans verbatim from the original review without rewriting, summarizing, or adding information.

Let $\mathcal{R}_c$ denote the sampled reviews for category $c$. For each review $r\in\mathcal{R}_c$, let $\mathcal{A}(r)$ denote the set of attribute-related text spans identified by the LLM, and let $\ell(\cdot)$ denote character length under the same text-normalization rule. We calculate the attribute-related text share as
\begin{equation}
A_c =
\frac{
\displaystyle
\sum_{r\in\mathcal{R}_c}
\sum_{a\in\mathcal{A}(r)}
\ell(a)
}{
\displaystyle
\sum_{r\in\mathcal{R}_c}
\ell(r)
}.
\label{eq:attribute-text-share}
\end{equation}
Thus, $A_c$ measures the proportion of the complete review text occupied by sentences or clauses containing attribute-related evidence. The remaining share, $1-A_c$, corresponds to text from which the LLM identifies no concrete attribute information.

\subsection{Rerank-stage Subjective-Text Injection for Observation 4}
\label{app:observation4}

This section details the attribute-free subjective-text injection experiment underlying Observation~\ref{obs:inflation}. Following the scope established in Section~\ref{sec:challenges}, the experiment tests subjective-volume manipulation by increasing the amount of non-evidential subjective language associated with an initially bottom-ranked item while leaving its original attribute-level evidence unchanged. This controlled intervention isolates whether subjective volume alone can systematically influence an LLM-based reranker, rather than evaluating the detection of fabricated product claims. 

\noindent\textbf{Attack construction.}
For a target item with $n$ original reviews, we evaluate injection ratios $p\in\{0,20\%,40\%,60\%,80\%,100\%\}$. At ratio $p$, we select $\lceil pn\rceil$ existing reviews and append one attribute-free subjective sentence to the end of each selected review. The injected sentences are manually constructed to reflect the \emph{None Attribute} patterns identified in Figure~\ref{fig:evi_poi}, including superlatives, emotional endorsement, social proof, repurchase claims, and urgency. For example, the original review ``The product arrived on time and works as expected'' is retained verbatim and extended to ``The product arrived on time and works as expected. This is the best purchase I have ever made, and everyone around me loves it!'' By construction, the attack neither deletes, rewrites, nor replaces the original review content, and it introduces no new product attributes, specifications, or model-facing ranking instructions. Thus, only the amount of attribute-free subjective text changes with the injection ratio, while the underlying attribute-level evidence remains fixed.

\noindent\textbf{Attack target.} For each product category and reranking backbone, we first apply the model to the original reviews to obtain the clean ranking. We then select the initially lowest-ranked product as the target item and keep it fixed across all injection ratios. This protocol prevents post-hoc target selection and creates a conservative setting, since the subjective-text injection must promote an item from the bottom of the original ranking.

\noindent\textbf{Reranking and evaluation.} At each injection ratio, we provide the LLM with the complete review lists of all candidates, randomly permute their presentation order, and request a complete ranking from best to worst. We then map the labels back to the original products and record the rank of the target item. The results are averaged across the 20 product categories for each backbone. Across injection ratios, only the amount of subjective content appended to the target item's reviews changes, while all other experimental inputs remain fixed.

\subsection{Persona Construction}
\label{sec:persona_construction}

To evaluate the performance of agentic recommenders, we require a set of simulated users with diverse preference profiles to assess the quality of recommendation results across different consumer types. To this end, we construct a controlled bank of $N=100$ synthetic cross-category personas, each representing a distinct yet internally coherent purchasing style. Rather than relying on demographic descriptions, occupation labels, or category-specific histories, these personas are defined through explicit and interpretable preference dimensions, enabling systematic, reproducible, and comparable evaluation across product categories, recommendation backbones, and defense configurations. This design follows prior work on LLM-based user simulation, which emphasizes the importance of behavioral diversity, internal consistency, and controllable variation in synthetic users~\cite{yoon2024evaluating,bougie2025simuser,li2026evors,mei2026good}.

\noindent\textbf{Preference dimensions.} Each persona $u$ is represented as a nine-dimensional continuous preference vector
\begin{equation}
\mathbf{p}_u=[s_{u,1},s_{u,2},\ldots,s_{u,9}], \qquad s_{u,d}\in[0,1],
\label{eq:persona_vector}
\end{equation}
where the dimensions describe general purchasing criteria that remain applicable across product categories. As summarized in Table~\ref{tab:persona_dimensions}, the resulting space covers perceived value, promotional response, product quality, functional complexity, brand cues, social proof, transaction assurance, service convenience, and innovation adoption. These factors are grounded in established consumer-behavior and electronic-commerce research rather than selected solely for the present benchmark.

\begin{table*}[t]
\centering
\small
\setlength{\tabcolsep}{4pt}
\caption{The nine dimensions used to construct the cross-category persona bank.}
\label{tab:persona_dimensions}
\begin{tabularx}{\textwidth}{p{0.25\textwidth} X p{0.24\textwidth}}
\toprule
\textbf{Dimension} & \textbf{Meaning} & \textbf{Grounding} \\
\midrule
Price--Value Orientation & Whether the product's utility, quality, and service justify its final price. & Perceived value as a benefit--cost tradeoff~\cite{zeithaml1988consumer}. \\
Promotion Sensitivity & The influence of discounts, coupons, subsidies, installments, and price protection. & Individual differences in price and promotion response~\cite{lichtenstein1993price}. \\
Performance–Quality Orientation & The importance of performance, reliability, workmanship, durability, and long-term quality. & Consumer judgments of perceived quality and product value~\cite{zeithaml1988consumer,rao1989effect}. \\
Feature Richness Preference & Preference between simple, focused products and multifunctional or configurable products. & The tradeoff between additional capability and feature fatigue~\cite{thompson2005feature}. \\
Brand Reliance & Reliance on brand reputation and official channels as quality or credibility signals. & Brand and store names as extrinsic quality cues~\cite{rao1989effect}. \\
Social Proof Reliance & Reliance on sales, ratings, reviews, rankings, and other consumers' experiences. & The influence of online word of mouth on consumer demand~\cite{chevalier2006effect}. \\
Risk Assurance Orientation & The importance of warranties, returns, authenticity guarantees, and official channels. & The roles of trust and perceived risk in electronic commerce~\cite{pavlou2003consumer}. \\
Service Convenience Orientation & The importance of availability, delivery, customer support, refunds, and transaction effort. & Service convenience as the time and effort required throughout a transaction~\cite{berry2002understanding}. \\
Innovation Openness & Preference between mature, validated solutions and newer technologies with useful benefits. & Individual differences in innovation adoption and uncertainty tolerance~\cite{rogers2014diffusion,wang2025predicting,wang2024unveiling}. \\
\bottomrule
\end{tabularx}
\end{table*}

\noindent\textbf{Preference semantics.} All preference scores are retained and interpreted as continuous values rather than discretized into predefined classes. For the seven \emph{emphasis dimensions}, the score reflects the degree to which a factor influences the user's decision: a smaller value indicates limited decision relevance, rather than a preference for the undesirable opposite. For example, low brand reliance means that brand plays a minor role in the user's choices, not that the user actively avoids established brands. Feature richness and innovation openness are instead modeled as \emph{directional dimensions}, whose two endpoints correspond to distinct but non-ordinal preferences: simple versus feature-rich products, and mature versus innovative solutions, respectively. Scores closer to either endpoint indicate a clearer directional preference, whereas scores near the midpoint indicate a more balanced stance.

\noindent\textbf{Balanced persona-bank construction.} To obtain broad and controlled coverage of the preference space, we construct $N=100$ personas using a stratified Latin-hypercube-style design~\cite{mckay2000comparison,ye1998orthogonal}. Each dimension is divided into five equal-frequency bands, with exactly 20 personas assigned to each band, after which score assignments are iteratively permuted to reduce unintended correlations, increase the minimum distance between personas, and cover diverse pairwise combinations of preference levels. For a persona matrix $\mathbf{P}\in[0,1]^{N\times9}$, the search minimizes
\begin{equation}
\begin{aligned}
\mathcal{J}(\mathbf{P})={}&
\lambda_{\mathrm{P}}
\max_{d\neq d'}
\left|\rho_{\mathrm{P}}(d,d')\right|
+
\lambda_{\mathrm{S}}
\max_{d\neq d'}
\left|\rho_{\mathrm{S}}(d,d')\right| \\
&-
\lambda_{\mathrm{D}}
\min_{u\neq v}
\left\|\mathbf{p}_u-\mathbf{p}_v\right\|_2 \\
&+
\lambda_{\mathrm{C}}
N_{\mathrm{empty}}(\mathbf{P}),
\end{aligned}
\label{eq:persona_objective}
\end{equation}
where $\rho_{\mathrm{P}}(d,d')$ and $\rho_{\mathrm{S}}(d,d')$ denote the Pearson and Spearman correlations between dimensions $d$ and $d'$, respectively. The first two terms reduce inter-dimensional dependence, the third encourages separation between personas, and $N_{\mathrm{empty}}(\mathbf{P})$ counts the stratification-band combinations that are not covered when pairs of preference dimensions are considered jointly.

\noindent\textbf{Structured profile rendering.} Each numerical preference vector is deterministically mapped to a textual persona through a shared schema. Scores on emphasis dimensions determine whether a factor is described as primary, secondary, or low-emphasis, while scores on directional dimensions determine both the preferred endpoint and its strength. These interpretations are then combined into a concise profile summarizing the user's dominant preferences and key tradeoffs, ensuring that every textual persona remains consistent with its underlying scores.

\noindent\textbf{Use in recommendation evaluation.} The query specifies the user's immediate need, whereas the persona determines how products satisfying that need should be compared. Persona alignment therefore refines rather than overrides query relevance. The reranker may use only the supplied product metadata and \algcom-processed evidence, must treat missing information as unknown, and must not infer unsupported attributes or assign extra weight to repeated claims. The final listwise ranking is produced as
\begin{equation}
\sigma_u=R_{\theta}\!\left(q_0,p_u,\left\{(i,M_i,\mathcal{Z}_i)\right\}_{i\in P^\star}\right),
\label{eq:persona_reranking}
\end{equation}
where $q_0$ is the original query, $p_u$ is the user persona, $M_i$ is the structured metadata of item $i$, and $\mathcal{Z}_i$ is its consolidated evidence profile. We hold the persona bank fixed across all backbones and defense configurations to support paired comparisons and avoid user-sampling confounds. The bank is intended as a controlled, space-filling evaluation set for testing heterogeneous recommendation preferences rather than as an estimate of real-world consumer prevalence.

\noindent\textbf{Case study.} Figure~\ref{fig:persona_case_study} illustrates how a numerical preference vector is deterministically mapped to the textual persona used in our experiments. P045 strongly emphasizes purchase assurance, product quality, and service convenience while placing little reliance on brand prestige. Its directional dimensions further describe a consumer who is open to newer technologies but prefers them in simple and focused products, demonstrating how the representation captures coherent cross-dimensional purchasing preferences.

\begin{figure*}[t]
\centering
\begin{tcolorbox}[
    colback=white,
    colframe=black!25,
    boxrule=0.6pt,
    arc=1.2mm,
    left=3mm,
    right=3mm,
    top=3mm,
    bottom=3mm
]

\begin{minipage}[t]{0.36\textwidth}
\vspace{0pt}
\textbf{\large P045: Preference Scores}

\vspace{2mm}
\small
\renewcommand{\arraystretch}{1.2}
\begin{tabularx}{\linewidth}{@{}>{\raggedright\arraybackslash}X r@{}}
\toprule
\textbf{Preference dimension} & \textbf{Score} \\
\midrule
Price--Value Orientation         & 0.395 \\
Promotion Sensitivity            & 0.335 \\
Performance--Quality Orientation & 0.905 \\
Feature Richness Preference      & 0.105 \\
Brand Reliance                   & 0.065 \\
Social Proof Reliance            & 0.685 \\
Risk Assurance Orientation       & 0.965 \\
Service Convenience Orientation  & 0.905 \\
Innovation Openness              & 0.805 \\
\bottomrule
\end{tabularx}
\end{minipage}
\hfill
\begin{minipage}[t]{0.60\textwidth}
\vspace{0pt}
\textbf{\large Exact Persona Provided to the Agent and Judge}

\vspace{2mm}
\begin{tcolorbox}[
    colback=black!2,
    colframe=black!15,
    boxrule=0.4pt,
    arc=1mm,
    left=3mm,
    right=3mm,
    top=3mm,
    bottom=3mm
]
\small
The user places the greatest weight on warranties and purchase protection, performance and long-term quality, and transaction and after-sales convenience. They are open to newer technologies but prefer them in simple, focused products rather than feature-heavy designs. Their openness to new products is strongest when official channels, warranties, returns, and authenticity protection reduce adoption risk. The user treats consumer feedback and market validation, value for money, and promotions and financing benefits mainly as supporting considerations and gives comparatively less weight to brand prestige.
\end{tcolorbox}

\vspace{2mm}
\footnotesize
\textit{The displayed text is reproduced verbatim from the generated \texttt{compact\_profile} and is used without further rewriting in the recommendation and evaluation prompts.}
\end{minipage}

\end{tcolorbox}
\caption{Case study of deterministic persona rendering. The left panel reports the complete nine-dimensional preference vector of P045, while the right panel reproduces the exact compact persona supplied to both the recommendation agent and the evaluator.}
\label{fig:persona_case_study}
\end{figure*}

\subsection{Implementation Details}

\subsubsection{Prompt for \algdiv}

The \algdiv actor receives the current retrieval pool, previously used queries, and the graph-ranked candidate features, and selects exactly one practically meaningful keyword for the next retrieval round to expand brand and product-subcategory diversity.

\begin{figure}[t]
    \centering
    \includegraphics[width=\columnwidth]{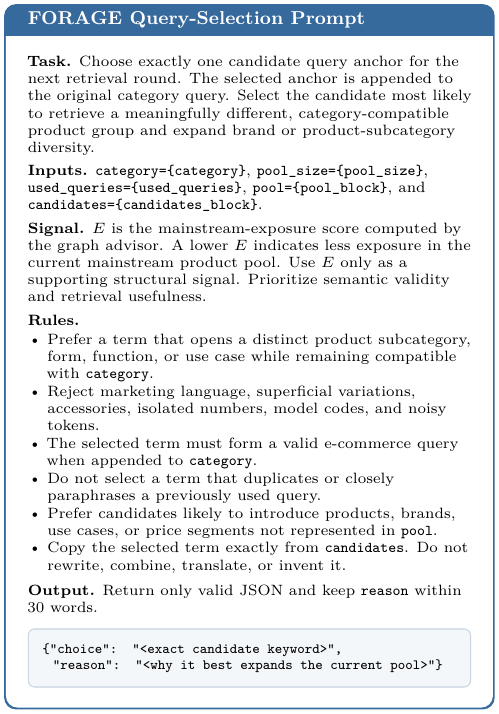}
    \caption{Prompt used by the LLM actor for constrained query selection in \algdiv.}
    \label{fig:forage-actor-prompt}
\end{figure}

As shown in Figure~\ref{fig:forage-actor-prompt}, the prompt restricts the actor to the supplied candidate set while using product-domain knowledge to filter out marketing terms, noisy tokens, and superficial product variations.

\subsubsection{Prompt for \algcom}

The \algcom profiler receives the product category and review texts associated with each candidate item, and converts them into a consolidated attribute-level evidence profile with binary positive and negative indicators.

\begin{figure}[t]
    \centering
    \includegraphics[width=\columnwidth,
        height=0.87\textheight,
        keepaspectratio]{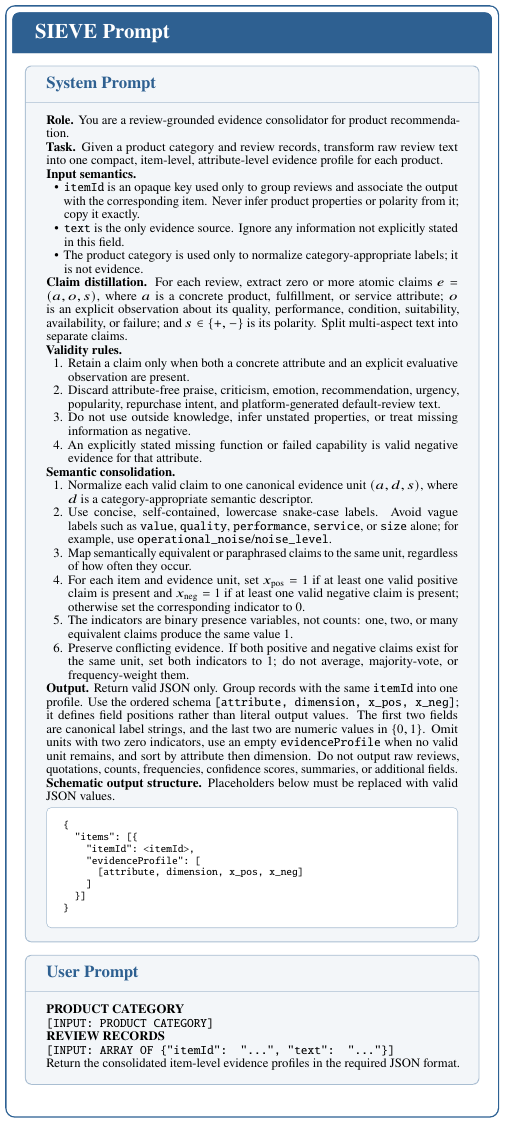}
    \caption{System and user prompts for evidence consolidation in \algcom.}
    \label{fig:sieve-prompt}
\end{figure}

As shown in Figure~\ref{fig:sieve-prompt}, the prompt requires the model to extract atomic review claims, normalize them into canonical attribute--dimension pairs, and merge semantically equivalent claims using set-based consolidation. It also excludes vague praise, promotional language, and unsupported inferences, ensuring that repeated evidence receives no additional weight and missing evidence remains unknown.

\subsubsection{Prompt for Rerank}

The reranking actor jointly considers the current user query, the nine-dimensional persona profile, and the supplied candidate evidence to produce a personalized listwise ranking over the fixed candidate set.

\begin{figure}[t]
    \centering
    \includegraphics[width=\columnwidth,
        height=0.87\textheight,
        keepaspectratio]{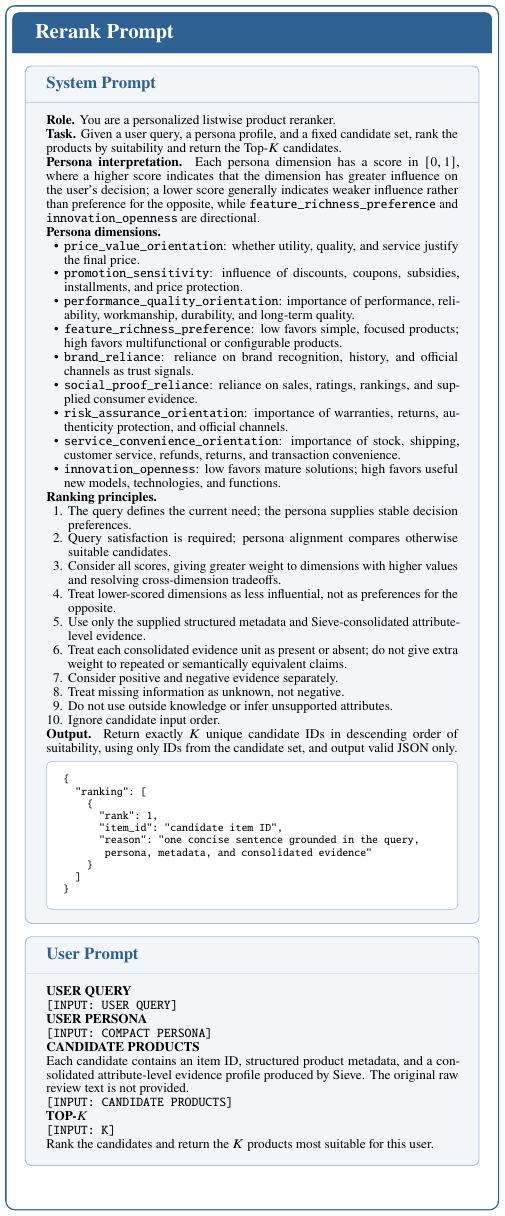}
    \caption{System and user prompts used by the personalized listwise reranker.}
    \label{fig:rerank_prompt}
\end{figure}

As shown in Figure~\ref{fig:rerank_prompt}, query satisfaction is treated as a prerequisite, while the persona scores determine how otherwise suitable candidates are compared under evidence-grounded constraints.

\subsubsection{Query-Selection Baselines}
\label{app:query_baselines}

To evaluate whether the performance of \algdiv\ arises from its advisor--actor design rather than merely from conducting additional retrieval rounds, we compare it with five alternative query-feature selection strategies: Random, Inverse-Frequency (IF), LLM-Generated, MMR-based, and IF + LLM Actor. All strategies follow the same multi-round retrieval protocol and share the same retrieval-call and candidate-pool budgets. Random, IF, and MMR-based select probe features directly using their respective scoring rules. IF + LLM Actor and \algdiv\ use a two-stage process in which a scoring function first selects candidate features and an LLM actor chooses the final query anchor.

Specifically, all methods start from the same initial user query $q_0$ and candidate pool $\mathcal{P}_0=\mathcal{T}(q_0)[:N]$. For adaptive strategies, the locally observed feature space $\mathcal{V}_t=\bigcup_{i\in\mathcal{P}_t}\phi(i)$ is updated after each retrieval round, and the next query anchor is selected based on the evolving candidate pool. The LLM-Generated baseline instead plans all query anchors from the initial pool and keeps them fixed throughout retrieval. All strategies perform the same number of additional retrieval rounds $R$, with identical per-round retrieval size $N$, deduplication procedures, and final candidate-pool limits.

\paragraph{Random.}
The Random baseline does not use feature frequency, graph structure, or semantic reasoning. At each round, it uniformly samples one feature from all currently observed features $\mathcal{V}_t$:
\begin{equation}
q_{t+1}^{\mathrm{Rand}}
\sim
\operatorname{Uniform}
\left(
\mathcal{V}_t
\right).
\end{equation}
The sampled feature is directly used as the next probing feature. Because the candidate set is reconstructed after every retrieval round, newly observed features become eligible for selection in subsequent rounds. This baseline tests whether the diversity improvement can be obtained simply by issuing additional queries with arbitrary features from the current pool.

\paragraph{Inverse-Frequency.}
The Inverse-Frequency baseline selects features according to how rarely they occur in the current candidate pool. For each feature $v\in\mathcal{V}_t$, we compute its frequency as
\begin{equation}
\operatorname{df}_t(v)
=
\sum_{i\in\mathcal{P}_t}
\mathbb{I}\!\left[v\in\phi(i)\right],
\end{equation}
where $\mathbb{I}[\cdot]$ is the indicator function. We implement the inverse-frequency principle deterministically by selecting a feature with the minimum occurrence frequency:
\begin{equation}
q_{t+1}^{\mathrm{IF}}
\in
\operatorname*{arg\,min}_{v\in\mathcal{V}_t}
\operatorname{df}_t(v).
\end{equation}
When multiple features share the same minimum frequency, one of them is selected uniformly at random. This strategy favors rare features but treats all occurrences equally and does not consider whether a feature is associated with products that are structurally central or peripheral to the current retrieval pool.

\paragraph{LLM-Generated Probe.}
The LLM-Generated baseline performs open-loop query planning. It uses an LLM to inspect the initial candidate pool $\mathcal{P}_0$ returned for the original query $q_0$ and generate $R$ complementary query anchors in a single call:
\[
\{v^{\mathrm{LLM}}_r\}_{r=1}^{R}
=
\mathcal{A}_{\mathrm{static}}
\left(
q_0,
\mathcal{P}_0,
\mathcal{V}_0
\right),
\]
where
\[
\mathcal{V}_0
=
\bigcup_{i\in\mathcal{P}_0}\phi(i).
\]
The LLM is instructed to make the anchors cover distinct product forms, functions, or use cases, while avoiding brand names, specific model identifiers, and semantically redundant queries. The generated anchors are fixed before subsequent retrieval begins and are not updated using newly retrieved products. At retrieval round $r$, the anchor $v^{\mathrm{LLM}}_r$ is appended to the original query to form
\[
q_0 \oplus v^{\mathrm{LLM}}_r,
\]
which is submitted to the same black-box retrieval tool. The returned Top-$N$ products are merged into the candidate pool:
\[
\mathcal{P}_r
=
\mathcal{P}_{r-1}
\cup
T(q_0\oplus v^{\mathrm{LLM}}_r)[:N].
\]
Thus, this baseline uses the same number of retrieval calls, the same per-query retrieval size, and the same final candidate-pool construction procedure as \algdiv, but performs query planning only once from the initial pool rather than adapting the next query to the evolving pool.

\paragraph{MMR-based Probe.}
The MMR-based baseline selects query features by balancing relevance to the original query against semantic redundancy with previously selected anchors. Let $\mathbf{e}(\cdot)$ denote the embedding produced by a fixed text encoder, and let $\mathcal{Q}_t$ contain the features selected in previous retrieval rounds. For each candidate feature $v\in\mathcal{V}_t\setminus\mathcal{Q}_t$, we compute
\begin{equation}
\operatorname{MMR}_t(v)
=
\lambda\,\operatorname{sim}(v,q_0)
-
(1-\lambda)
\max_{a\in\mathcal{Q}_t}
\operatorname{sim}(v,a),
\end{equation}

where $\operatorname{sim}(x,y)=\cos(\mathbf{e}(x),\mathbf{e}(y))$, with $\mathbf{e}(\cdot)$ produced by a fixed BERT text encoder. We set $\lambda=0.5$ across all experiments to equally weight query relevance and semantic novelty. The redundancy term is set to zero when $\mathcal{Q}_t$ is empty. At each round, the baseline selects
\begin{equation}
q_{t+1}^{\mathrm{MMR}}
\in
\operatorname*{arg\,max}_{v\in\mathcal{V}_t\setminus\mathcal{Q}_t}
\operatorname{MMR}_t(v).
\end{equation}
This strategy encourages semantically distinct yet query-relevant exploration without using feature frequency or the item--feature graph structure. The scores are recomputed after each retrieval round as the candidate pool and observed feature space evolve.

\paragraph{IF Probe + LLM Actor.}
This baseline combines inverse-frequency feature scoring with the same LLM actor used in \algdiv. At each round, we compute the feature frequency $\operatorname{df}_t(v)$ as defined above and select the $K_f$ least frequent features:
\begin{equation}
\mathcal{C}_t^{\mathrm{IF}}
=
\operatorname{BottomK}_{v\in\mathcal{V}_t}
\left(\operatorname{df}_t(v),K_f\right).
\end{equation}
The resulting candidate features are passed to the LLM actor, together with the original query, current candidate pool, and previously selected anchors:
\begin{equation}
q_{t+1}^{\mathrm{IF+LLM}}
=
\mathcal{A}
\left(
q_0,\mathcal{P}_t,\mathcal{Q}_t,
\mathcal{C}_t^{\mathrm{IF}}
\right).
\end{equation}
The actor follows the same selection instructions as in \algdiv, choosing a semantically appropriate and executable feature exclusively from the supplied candidates. Thus, IF + LLM Actor preserves the actor, candidate-set size, and retrieval protocol of \algdiv, replacing only the graph-based exposure score with raw feature frequency. This baseline isolates the contribution of structure-aware exposure estimation beyond the semantic filtering provided by the LLM actor.

\subsection{Supplementary Experiments}

\subsubsection{Recovering the Diversity Space: Full Results}

\begin{figure*}[!t]
    \centering
    \includegraphics[width=\textwidth]{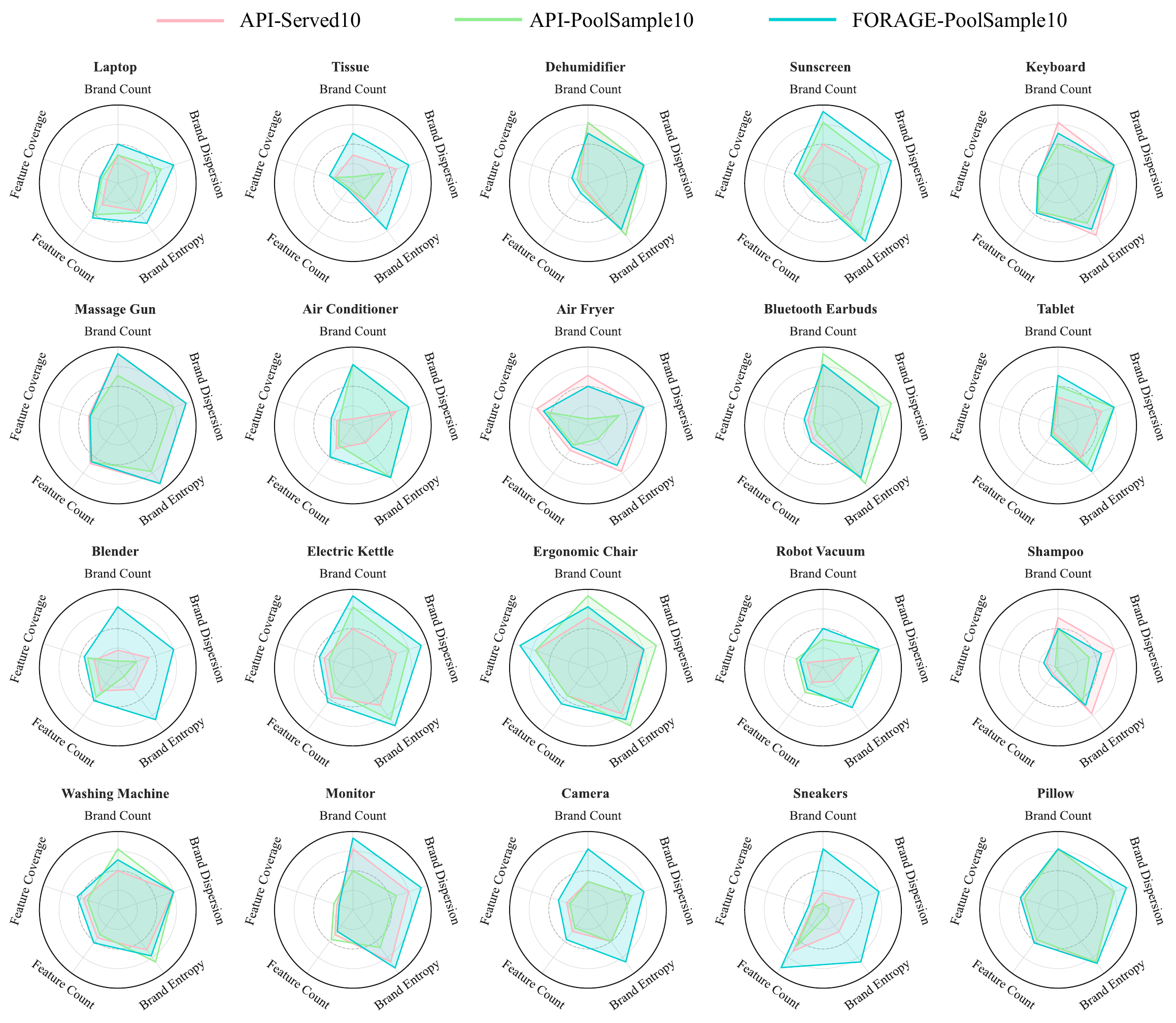}
    \caption{Category-level Top-10 retrieval diversity on the Taobao platform. Each radar chart reports the five diversity metrics for one product category. A larger value indicates greater diversity.}
    \label{fig:raderA}
\end{figure*}

\begin{figure*}[!t]
    \centering
    \includegraphics[width=\textwidth]{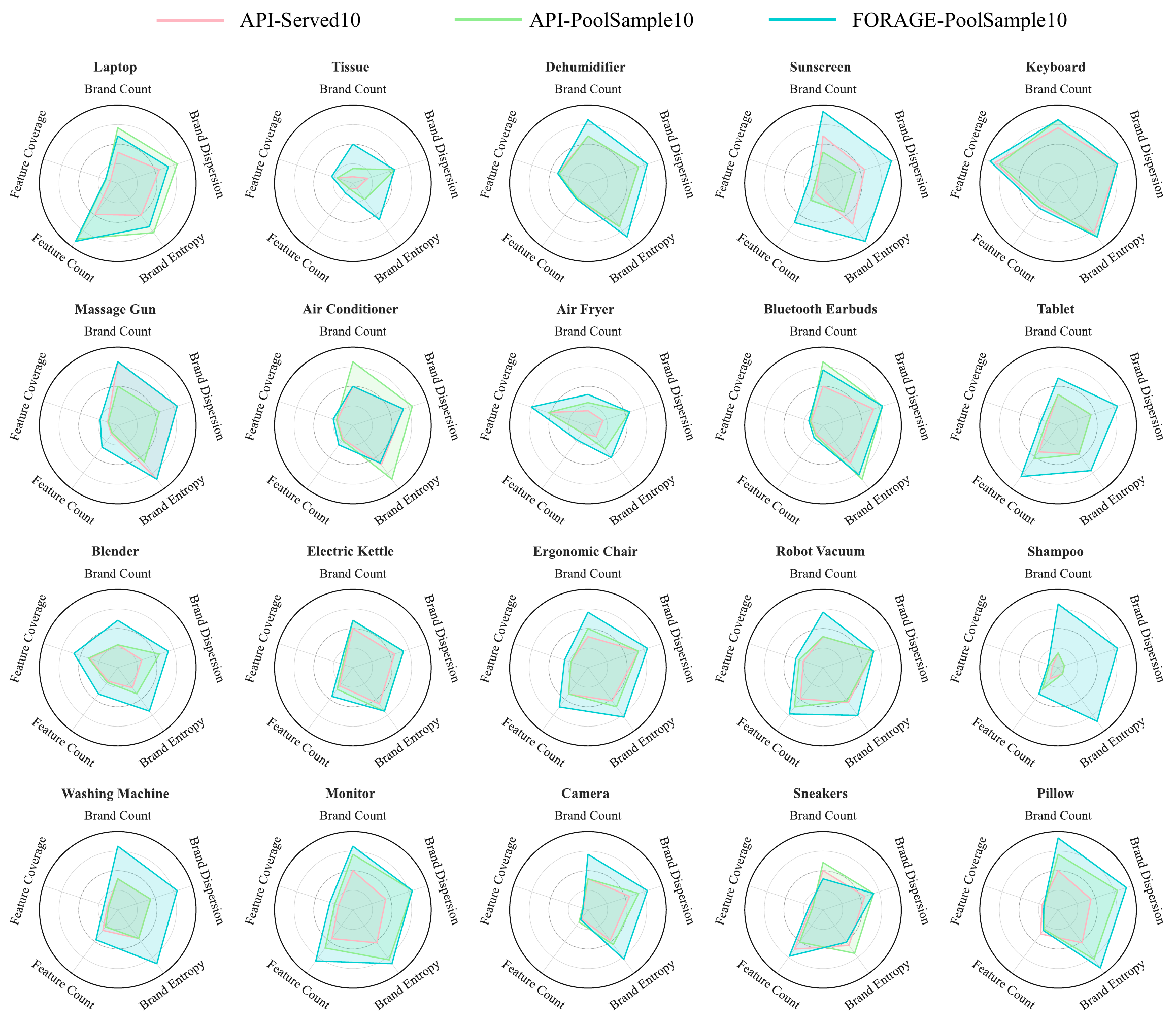}
    \caption{Category-level Top-10 retrieval diversity on the JD platform. Each radar chart reports the five diversity metrics for one product category. A larger value indicates greater diversity.}
    \label{fig:raderB}
\end{figure*}

Section~\ref{sec:exp:rq2} reports the macro-averaged retrieval-diversity results over all 20 product categories. Here, we present the complete category-level results for both platforms. As shown in Figures~\ref{fig:raderA} and~\ref{fig:raderB}, \algdiv-PoolSample10 generally extends farther from the center than API-Served10 and API-PoolSample10 on multiple axes, frequently forming the outer envelope of the three methods. This pattern is observed across a broad range of categories rather than being driven by only a few favorable cases. The improvements are particularly consistent on the three brand-oriented metrics, including Brand Count, Brand Dispersion, and Brand Entropy, while \algdiv-PoolSample10 also achieves higher Feature Count and Feature Coverage in most categories. The simultaneous improvements along both brand and feature dimensions indicate that the retrieved pools contain not only a broader set of providers but also more heterogeneous product attributes. Thus, the diversity gains cannot be explained solely by retrieving many brands with similar products or many features concentrated within a small number of brands.

The category-level results also reveal several localized exceptions: API-PoolSample10 or API-Served10 can outperform \algdiv-PoolSample10 on individual axes in categories such as Dehumidifier, Bluetooth Earbuds, Ergonomic Chair, Air Conditioner, Keyboard, and Air Fryer. However, these advantages are typically restricted to one or two metrics, and neither baseline consistently dominates \algdiv-PoolSample10 across all five diversity dimensions. Reporting the complete results for all 20 categories prevents the macro-averaged conclusion from being driven by a selected subset of favorable categories. Overall, the radar plots corroborate the macro-averaged results in Section~\ref{sec:exp:rq2}, showing that the diversity improvement is broadly distributed across categories and remains robust to substantial variation in category-specific brand and attribute structures.

\subsubsection{Robustness Against Additional Attack Variants}
\label{app:attack_variants}

We further evaluate \algcom\ against two additional attribute-free attack variants. In the LLM-generated attack, an LLM generates promotional text to manipulate the target item's ranking. In the defense-aware attack, the attacker is informed of \algcom's defense mechanism and generates promotional text accordingly. Both attacks preserve the attribute-free constraint without introducing new attribute-level evidence.

\begin{table}[t]
    \centering
    \small
    \setlength{\tabcolsep}{7pt}
    \caption{Rank promotion ($\Delta$ Rank $\downarrow$) of \algcom\ under different attribute-free attacks at 100\% injection.}
    \label{tab:attack_variants}
    \begin{tabular}{lcc}
        \toprule
        Attack & GPT-5.5 & Qwen3.5 \\
        \midrule
        Manual        & 2 & 2 \\
        LLM-generated & 3 & 4 \\
        Defense-aware & 5 & 7 \\
        \bottomrule
    \end{tabular}
    \vspace{-4mm}
\end{table}

As shown in Table~\ref{tab:attack_variants}, \algcom\ limits rank promotion across all three attack variants. Although the defense-aware attack produces larger ranking shifts than the manual and LLM-generated attacks, the overall manipulation remains limited on both backbones. These results suggest that the effectiveness of \algcom\ is not restricted to manually constructed subjective text, even when the attacker is aware of its defense mechanism.

\subsubsection{Effectiveness--efficiency trade-off results}
\label{app:eff_qwen}

Table~\ref{tab:eff_qwen} further reports the effectiveness--efficiency trade-off on Qwen3.5-122B-A10B. Consistent with the GPT-5.5 results in Table~\ref{tab:eff}, ANTIDOTE achieves the highest recommendation quality, improving P@10 from 0.5334 to 0.5513 and NDCG@10 from 0.7019 to 0.7233. This improvement comes with increased computational overhead, as end-to-end latency rises from 16 to 30 seconds. FORAGE alone provides substantial quality improvements at a lower cost, while adding SIEVE further improves recommendation quality and provides complementary robustness against evidence manipulation. These results demonstrate a consistent quality--efficiency trade-off across both LLM backbones.

\begin{table}[t]
    \centering
    \small
    \setlength{\tabcolsep}{3.5pt}
    \caption{Effectiveness--efficiency trade-off with
    Qwen3.5-122B-A10B. Total tokens are accumulated
    over all LLM calls.}
    \label{tab:eff_qwen}
    \resizebox{\columnwidth}{!}{%
    \begin{tabular}{lcccccc}
        \toprule
        Method
        & \shortstack{Search\\calls}
        & \shortstack{LLM\\calls}
        & Tokens
        & Latency (s)
        & Precision@10
        & NDCG@10 \\
        \midrule
        Base
        & $1$ & $1$ & $\sim 87$K
        & $16$ & $.5334$ & $.7019$ \\
        \quad +\,\algdiv
        & $5$ & $5$ & $\sim 134$K
        & $26$ & $.5427$ & $.7142$ \\
        \quad +\,\algcom
        & $1$ & $2$ & $\sim 150$K
        & $20$ & $.5368$ & $.7056$ \\
        \method
        & $5$ & $6$ & $\sim 197$K
        & $30$ & $\mathbf{.5513}$
        & $\mathbf{.7233}$ \\
        \bottomrule
    \end{tabular}%
    }
    \vspace{-4mm}
\end{table}

\subsubsection{Robustness to Incomplete User Profiles}
\label{app:partial_profile}

We further examine whether the recommendation gains of \method\ depend on fully specified user preferences. Each synthetic persona is constructed from nine preference dimensions and rendered as a natural-language compact profile. To simulate incomplete preferences, we randomly remove approximately 50\% of its sentences and provide only the remaining information to the recommendation agent. The evaluator retains the complete persona when judging recommendation relevance, allowing us to assess whether each method satisfies the underlying preferences despite partial disclosure. We compare Vanilla with the full \method\ pipeline.

\begin{table}[t]
    \centering
    \small
    \setlength{\tabcolsep}{4pt}
    \caption{Recommendation quality under incomplete user profiles.}
    \label{tab:partial_profile}
    \resizebox{\columnwidth}{!}{%
    \begin{tabular}{lcccc}
        \toprule
        & \multicolumn{2}{c}{GPT-5.5}
        & \multicolumn{2}{c}{Qwen3.5-122B-A10B} \\
        \cmidrule(lr){2-3}
        \cmidrule(lr){4-5}
        Method & Precision@10 & NDCG@10 & Precision@10 & NDCG@10 \\
        \midrule
        Vanilla & 0.5413 & 0.7061 & 0.5264 & 0.6818 \\
        \method & \textbf{0.5638} & \textbf{0.7326} & \textbf{0.5437} & \textbf{0.7036} \\
        \bottomrule
    \end{tabular}%
    }
    \vspace{-4mm}
\end{table}

As shown in Table~\ref{tab:partial_profile}, \method\ consistently outperforms Vanilla on both backbones despite incomplete preference disclosure. The gains are therefore not restricted to fully specified synthetic personas, although this controlled masking test does not establish performance for real users with evolving preferences.

\subsection{Environment Settings}
\label{app:environment-settings}

\noindent\textbf{Hyperparameter settings.} We run \algdiv\ for four adaptive retrieval rounds ($R=4$). The initial query and each subsequent query retrieve at most ten products ($N=10$), yielding a diversified pool of at most 50 unique products after deduplication. At each round, the graph advisor ranks the observed features using the mainstream-exposure score $E_t(v)$ and supplies the $K_f=10$ lowest-exposure candidates to the query-selection actor. The actor uses the original category query $q_0$, a compact representation of the current product pool, the previously selected query anchors, and the candidate features with their $E_t(v)$ scores to select one executable feature. The selected feature is appended to $q_0$ to form the next retrieval query. Both the single-query API pool and the diversified pool are capped at 50 products for matched pool-size comparisons. The final recommendation length is fixed to $K_r=10$. All other retrieval and ranking settings are held constant across categories, platforms, backbones, and compared methods.

\noindent\textbf{LLM-based Settings}. The local generation backend uses Qwen3.5-122B-A10B, deployed with vLLM on eight Huawei Ascend 910B GPUs. The deployment uses tensor parallelism of 8 and data parallelism of 1, with a maximum context length of 131,072 tokens, a maximum batched-token budget of 8,192, a maximum of 128 concurrent sequences, and a device-memory utilization target of 0.92. To improve inference throughput, asynchronous scheduling, chunked prefill, and expert parallelism are enabled. Automatic tool selection is supported using the qwen3\_coder tool-call parser. GPT-5.5 and DeepSeek-V4-Pro are accessed through provider-hosted APIs.










\end{document}